\documentclass[12pt]{article}

\pdfoutput=1

\usepackage{tikz-cd}
\usetikzlibrary{arrows,snakes,shapes.arrows,decorations.markings}
     \tikzset{>=triangle 90}
     \tikzstyle{bbc}=[draw,circle,fill=black,scale=.75]
     \tikzstyle{rc}=[circle,fill=red,scale=.6]
     \tikzstyle{wc}=[draw,circle,scale=.75]

\usepackage[top=80pt,bottom=85pt,left=85pt,right=85pt]{geometry}
\usepackage{amssymb}
\usepackage{amsmath}
\usepackage{setspace}
\usepackage{sectsty}
\usepackage{cite}
\usepackage{bm}
\usepackage{bbm}
\usepackage[utf8]{inputenc}
\usepackage[T1]{fontenc}
\usepackage{comment}
\usepackage[english]{babel}
\usepackage{afterpage}
\usepackage{bbold}
\usepackage{mathdots}
\usepackage{physics}
\usepackage{makecell,booktabs}
\usepackage{multirow}
\usepackage{lscape}
\usepackage{mathtools}
\usepackage{dashrule}

\usepackage{graphicx,subcaption}
\usepackage{setspace}
\usepackage{float}
\usepackage{booktabs}
\usepackage[debug,pageanchor=false]{hyperref}
\definecolor{link}{rgb}{.8,.15,.1}
\definecolor{pigment}{rgb}{0.36, 0.54, 0.66}
\definecolor{pigment2}{rgb}{0.19, 0.55, 0.91}
\definecolor{pigment3}{rgb}{0.2, 0.2, 0.6}
\definecolor{light-gray}{gray}{0.75}
\hypersetup{colorlinks=true,linkcolor=link,citecolor=link,urlcolor=link,linktocpage}

\usepackage{array,multirow,booktabs,longtable}
\usepackage{mathrsfs}
\usepackage{tikz-cd} 
\usetikzlibrary{backgrounds, arrows,calc,shapes,decorations.pathreplacing, decorations.markings, automata,positioning}
\tikzset{%
  >={Latex[width=2mm,length=2mm]},
            base/.style = {rectangle, rounded corners, draw=black,
                           minimum width=4cm, minimum heigwht=1cm,
                           text centered, font=\sffamily},
  activityStarts/.style = {base, fill=orange!15},
       startstop/.style = {base, fill=orange!15},
    activityRuns/.style = {base, fill=orange!15},
         process/.style = {base, minimum width=2.5cm, fill=orange!15,
                           font=\ttfamily},
}

\newcommand{\red}[1]{}

\tikzset{
        cvertex/.style={circle,draw=black,inner sep=1pt,outer sep=3pt},
        vertex/.style={circle,fill=black,inner sep=1pt,outer sep=3pt},
        star/.style={circle,fill=yellow,inner sep=0.75pt,outer sep=0.75pt},
        tvertex/.style={inner sep=1pt,font=\scriptsize},
        gap/.style={inner sep=0.5pt,fill=white}}

\tikzstyle{mybox} = [draw=black, fill=blue!10, very thick,
    rectangle, rounded corners, inner sep=10pt, inner ysep=20pt]
\tikzstyle{boxtitle} =[fill=blue!50, text=white,rectangle,rounded corners]

\newcolumntype{C}{>{\hfil$}p{3cm}<{$\hfil}}
\newcolumntype{P}{>{\hfil$}p{7.7cm}<{$\hfil}}
\newcolumntype{L}{>{\hfil$}p{2.8cm}<{$\hfil}}
\newcolumntype{S}{>{\hfil$}p{1.8cm}<{$\hfil}}
\newcolumntype{R}{>{\hfil$}p{5.2cm}<{$\hfil}}
\newcolumntype{U}{>{\hfil$}p{4.2cm}<{$\hfil}}
\newcolumntype{Q}{>{\hfil$}p{6.4cm}<{$\hfil}}
\newcolumntype{T}{>{\hfil$}p{1.9cm}<{$\hfil}}
\newcolumntype{V}{>{\hfil$}p{5.8cm}<{$\hfil}}
\newcolumntype{H}{>{\hfil$}p{1.8cm}<{$\hfil}}
\newcolumntype{A}{>{\hfil$}p{6cm}<{$\hfil}}
\newcolumntype{B}{>{\hfil$}p{2cm}<{$\hfil}}
\newcolumntype{D}{>{\hfil$}p{7.4cm}<{$\hfil}}

\newcommand{\todo}[1]{}
\renewcommand{\todo}[1]{{\color{red} TODO: {#1}}}
\renewcommand{\red}[1]{{\color{red} {#1}}}

\newcommand{\be}{\begin{equation}}  
\newcommand{\ee}{\end{equation}}  
\newcommand{\bea}{\begin{align}}
\newcommand{\eea}{\end{align}}
\newcommand{\bp}{\begin{bmatrix*}[r]}  
\newcommand{\bpp}{\begin{bmatrix}}  
\newcommand{\epp}{\end{bmatrix}}  
\newcommand{\bcd}{\begin{center}
\begin{tikzcd}}
\newcommand{\ecd}{\end{tikzcd} \end{center}}
\newcommand{\bpm}{\begin{pmatrix}}  
\newcommand{\eem}{\end{pmatrix}}

\makeatletter
\@addtoreset{equation}{section}
\makeatother

\begin{document}

\begin{titlepage}

\begin{center}

\vskip .3in \noindent

{\Large \bf{D2-brane probes of non-toric cDV threefolds  \\ 

via monopole superpotentials

\vspace{.5cm} \hspace{1mm}    }}

\bigskip\bigskip\bigskip

Andr\'es Collinucci$^{a}$, Marina Moleti$^{b}$ and Roberto Valandro$^{c}$ \\

\bigskip


\bigskip
{\footnotesize
 \it

$^a$ \footnotesize Physique Th\'eorique et Math\'ematique and International Solvay Institutes\\
Universit\'e Libre de Bruxelles, C.P. 231, 1050 Brussels, Belgium \\
\vspace{.25cm}
$^b$ SISSA and INFN, Via Bonomea 265, I-34136 Trieste, Italy\\
\vspace{.25cm}
$^c$ Dipartimento di Fisica, Universit\`a di Trieste, Strada Costiera 11, I-34151 Trieste, Italy \\
and INFN, Sezione di Trieste, Via Valerio 2, I-34127 Trieste, Italy	
}

\vskip .7cm
{\scriptsize \tt collinucci dot phys at gmail dot com \hspace{1cm} 
    mmoleti at sissa dot it \\
   roberto dot valandro at ts dot infn dot it
    }

\vskip 2cm
 \end{center}

\begin{abstract}
We develop a framework to construct worldvolume gauge theories on D2-branes probing compound Du Val (cDV) Calabi–Yau threefold singularities. By viewing these singularities as ADE surface fibrations over the complex $w$-plane, we encode their geometry in a Higgs field $\Phi(w)$. A D2-brane probe perceives $\Phi(w)$ as an $\mathcal{N}=2$ deformation of its 3d $\mathcal{N}=4$ affine Dynkin quiver gauge theory via polynomial and monopole superpotential terms. By exploiting 3d mirror symmetry, we obtain an effective theory that correctly reproduces the quiver-collapsing mechanism known in the mathematical literature. We present several examples, including non-toric and non-resolvable cases.
\end{abstract}

\noindent

\vfill
\eject

\end{titlepage}

\tableofcontents

\newpage

\section{Introduction}

D-branes probing a singular Calabi-Yau geometry are described by supersymmetric quiver gauge theories whose vacuum moduli space reproduces the local geometry of the singularity. This correspondence between brane physics and singular geometry has been instrumental in understanding both the structure of supersymmetric gauge theories and the resolution of singularities.

For toric Calabi-Yau threefolds, a rich arsenal of techniques—including brane tilings, dimer models and toric diagrams \cite{Franco:2005rj,Franco:2005sm,Hanany:2005ie,Feng:2000mi}—provides explicit, algorithmic constructions of the probe brane theories.

Progress beyond the toric setting is substantially more limited \cite{Morrison:1998cs, Beasley:1999uz, Berenstein:2001nk, Wijnholt:2002qz, Aspinwall:2004jr, Aspinwall:2010mw}. Matrix-factorization techniques have provided probe theories for specific non-toric hypersurface singularities, including Reid's pagodas and Laufer-type examples \cite{Aspinwall:2010mw}. 

There is, however, no systematic method for constructing probe-brane theories of compound Du Val singularities (cDV). These are the threefold analogs of classical ADE surface singularities, and form a rich class of non-toric geometries that arise naturally in M-theory and F-theory compactifications \cite{Closset:2020scj,Aspinwall:2010mw, Collinucci:2018aho,Collinucci:2019fnh}, where they engineer superconformal field theories in various dimensions. While the geometric structure of cDV singularities has been extensively studied \cite{pagodas,pinkham,Karmazyn:2017aa,Donovan_2016}, a unified approach to constructing their probe-brane theories that goes beyond specific examples has been lacking.

In this paper, we accomplish that very goal. Our approach is algebraic rather than combinatorial: following the logic of \cite{Collinucci:2021wty,Collinucci:2021ofd,DeMarco:2021try,Collinucci:2022rii,DeMarco:2022dgh,Sangiovanni:2024nfz}, we introduce a complex Higgs field $\Phi(w)$ that characterizes the cDV fibration structure. This Higgs field, valued in the Lie algebra of the corresponding ADE group, efficiently encodes the threefold geometry: the fibration over the base $\mathbb{C}_w$, the resolution pattern of the ADE fiber, and the monodromy of vanishing cycles. 

Our method applies to both \emph{monodromic} and \emph{non-monodromic} fibrations (see Section \ref{subsec:higgsfield} for the definitions), and to resolvable and non-resolvable singularities. More precisely, we start from the three-dimensional (3d) $\mathcal{N}=4$ affine Dynkin quiver gauge theory on a D2-brane probing the corresponding ADE surface singularity, and we introduce a background Higgs field $\Phi(w)$ that encodes an $\mathcal{N}=2$ deformation of the probe theory, possibly involving monopole operators in the superpotential. In the class of examples treated in this paper 3d mirror symmetry allows us to rewrite the resulting theory in a standard infrared Lagrangian form, with polynomial superpotential interactions. Under renormalization-group (RG) flow, the quiver undergoes a collapse to a smaller effective quiver.
This collapse matches the quiver-collapsing mechanism appearing in Karmazyn's mathematical description of cDV threefolds \cite{Karmazyn:2017aa, VandenBergh:2004}.

In order to exploit 3d mirror-symmetry techniques, we study D2-brane probes of cDV singularities times a circle, rather than directly the four-dimensional $\mathcal N=1$ theories engineered by D3-brane probes. For a large class of cDV threefolds, the resulting three-dimensional $\mathcal{N}=2$ theory can be described as a deformation of a 3d $\mathcal{N}=4$ quiver gauge theory by monopole superpotential terms determined by the profile of $\Phi(w)$. 
These monopole deformations can be systematically analyzed (see \cite{Collinucci:2016hpz,Benini:2017dud,Collinucci:2017bwv}) using local 3d mirror symmetry \cite{Hanany:1996ie, Intriligator:1996ex,deBoer:1996mp, Kapustin:1999ha}, allowing us to trade monopole operators for polynomial interactions in elementary fields. The resulting effective theory admits a standard Lagrangian description, and its Higgs branch moduli space reproduces the cDV threefold. Upon decompactification (taking the circle radius to zero), this construction recovers the four-dimensional physics of D3-branes probing the singularity.

We verify our construction by reproducing known results, confirming agreement with existing techniques where available. Importantly, we successfully apply our method to examples beyond the reach of previous approaches:
1)~simple flops of length 2, representing a general non-toric family parametrized by arbitrary polynomial base changes;
2)~the non-resolvable $(A_2, D_4)$ threefold, where standard techniques based on resolutions fail.
These examples demonstrate that our approach provides a systematic framework to construct probe theories for cDV singularities beyond the toric and resolvable regime.

The paper is organized as follows. Section~\ref{Sec:D3vsD2} explains our strategy of working with D2-branes and extracting four-dimensional (4d) physics through decompactification. 
Section~\ref{Sec:3dTheories} reviews 3d supersymmetric field theories and monopole superpotentials. In Section~\ref{Sec:cDVHiggs}, we review the description of cDV threefolds in terms of a Higgs field $\Phi(w)$.
 Section~\ref{Sec:D2onADEfib} presents our main construction for non-monodromic and monodromic fibrations. Section~\ref{Sec:ReidPagodas}, Section~\ref{Sec:D4families} and Section~\ref{Sec:NonResolv} apply the method to examples. We conclude in Section~\ref{Sec:Conclusions} with future directions.

\section{D3-branes vs D2-branes probing CY threefolds}\label{Sec:D3vsD2}

In this paper we study three-dimensional theories on D2-branes probing cDV Calabi-Yau (CY) threefold singularities. These theories are interesting in their own right, but they are also useful because, in a suitable limit, they capture the geometric branch relevant to the four-dimensional theories engineered by D3-branes probing the same singularities. This provides a practical strategy to extract the geometry encoded by the probe theory while exploiting powerful three-dimensional tools, in particular local mirror symmetry and monopole superpotential deformations.

More concretely, start from a D3-brane probing a Calabi--Yau threefold singularity $X$. Compactifying one worldvolume direction on a circle and performing T-duality along it maps this configuration to a D2-brane probing $X\times S^1$. The resulting probe theory is a three-dimensional $\mathcal N=2$ theory, which should be viewed as the circle reduction of the four-dimensional D3-brane theory, possibly supplemented by genuinely three-dimensional effects.

The vacuum moduli space of 3d $\mathcal{N}=2$ theories is notoriously more complex than the 3d $\mathcal{N}=4$ case. In theories with only four supercharges, the distinction between the Coulomb Branch (CB) and the Higgs Branch (HB) is less rigid: such branches can be lifted by quantum corrections, or they may intersect in a way that prevents a clear-cut separation. Furthermore, the $U(1)_R$ symmetry, which typically helps define these branches, can mix with flavor symmetries along the RG flow. 

However, for theories possessing a well-defined UV Lagrangian, we can employ a standard working definition: the CB is parametrized by the scalar field $\sigma$ in the vector multiplet together with the dual photon $\gamma$, while the HB is parametrized by the VEVs of the matter fields. 

The physical interpretation of the 3d Coulomb branch scalar differs on the two sides of the T-duality. In Type IIB, before T-duality, the scalar $\sigma$ descends from the 4d photon compactified on the circle: it is the Wilson line $\oint A_4$ around the compact direction, and hence exhibits periodicity with period set by the radius $R$. After T-duality to Type IIA, the same scalar $\sigma$ parametrizes the position of the D2-brane along the dual circle of radius $r \propto \alpha'/R$. 

In the limit $r \to 0$ (corresponding to the 4d decompactification limit $R \to \infty$), these two descriptions converge to the same conclusion: the Coulomb branch is forced to collapse. On the IIB side, as $R \to \infty$, the Wilson line is frozen at zero. On the IIA side, as $r \to 0$, the dual circle shrinks and the D2-brane position loses its physical meaning, forcing $\langle \sigma \rangle = 0$. Thus, in both descriptions, the decompactification limit naturally restricts the 3d dynamics to the origin of the Coulomb branch.

Consequently, to recover the 4d D3-brane theory from the 3d D2-brane perspective, one must work at the origin of the Coulomb branch. This restriction isolates the Higgs branch, which remains robust under the compactification. Notably, the tree-level superpotential $W$, being a holomorphic function of the chiral matter fields, remains unaltered in this transition. 

In summary, the three-dimensional D2-brane theory provides a useful route to the geometric branch relevant for the four-dimensional D3-brane probe. The working assumption throughout this paper is that, after taking the decompactification limit and restricting to the origin of the Coulomb branch, the branch of vacua selected by the F-term equations reproduces the Calabi--Yau threefold probed by the D3-brane\footnote{We thank Amihay Hanany for pointing out to us the importance of understanding the interplay between the 3d and 4d theories, and how the former by itself isn't sufficient to fully isolate the CY geometry.}. In the class of examples considered below, this is precisely the branch captured by the effective $\mathcal N=2$ theory obtained from the D2-brane construction.
Algebraically, this corresponds to the space of solutions to the F-term equations derived from the superpotential,
\begin{equation}
    \mathcal{M}_{HB} = \{ \langle X_{i} \rangle \, | \, \partial_{X_{i}} W = 0 \} / / G \, , \nonumber
\end{equation}
where $G$ is the gauge group of the quiver. In all the examples analyzed in this paper, $\mathcal M_{\rm HB}$ agrees with the cDV threefold geometry.


\section{3d theories and monopole superpotentials}\label{Sec:3dTheories}

\subsection{Supersymmetric 3d gauge theories}

Before delving into our procedure, let us briefly explain the motivation for our indirect approach of using 3d quiver gauge theories to ultimately build 4d theories on D3-probes. Given the plethora of methods known since the early 2000's for D3-branes at singularities, this question is justified. 

First, we remind the reader that most of the techniques known in the literature, such as brane tilings, the fast inverse algorithms, exceptional collections (see \cite{Yamazaki:2008bt} and the multitude of references therein) are designed for and strictly limited to \emph{toric varieties}. In contrast, all of the cDV threefolds with isolated singularities (except for the conifold) are \emph{non-toric}, rendering all such methods powerless.

The abelian topological symmetry is enhanced to a non-abelian symmetry group corresponding to the ADE Lie algebra $\mathcal{G}$ \cite{Gaiotto:2008sa,Bashkirov:2010kz,Cremonesi:2013lqa}. In this enhancement, the topological $U(1)_T$ generators span the Cartan subalgebra of $\mathcal{G}$. 
More generally, whenever a quiver contains a set of balanced nodes, the associated non-abelian topological symmetry corresponds to the Lie algebra defined by the related Dynkin diagram. Furthermore, if one considers a balanced sub-quiver within a larger balanced quiver, the corresponding symmetry is given by the subalgebra associated with that subdiagram.

The monopole operators are labeled by their charges under the topological $U(1)^r$ symmetry. 
In $\mathcal{N}=4$ quiver gauge theories, these monopole operators reside in supermultiplets together with the current of the non-abelian symmetry. Together with the scalar fields $\varphi_i$, they can be assembled into an object transforming in the adjoint representation of the Lie algebra $\mathcal{G}$. This is the moment map $\mu$ for the topological symmetry.

\subsubsection*{Basic example}

\paragraph{Theory A:}
Let us consider a 3d $\mathcal{N}=2$ $U(1)$ gauge theory with $N+1$ flavors of positron/electron pairs $(Q_i, \tilde Q_i)$ and no superpotential. 
There is an $SU(N+1) \times SU(N+1)$ flavor symmetry that rotates the $Q_i$ and $\tilde Q_i$, respectively. The diagonal $SU(N+1)_\Delta$ has the moment map defined by mesons, transforming in the adjoint of $SU(N+1)_\Delta$:
\begin{equation}
    M_i^j = Q_i\,\tilde Q^j \:.
\end{equation} 
We can add mass deformations that break the full flavor group to its diagonal as follows:
\begin{equation}\label{eq:massdef}
\delta_{m} W = Q\cdot \Phi \cdot \tilde Q\,,
\end{equation} 
where $\Phi$ is a background, non-dynamical $SU(N+1)$-valued Higgs field. 

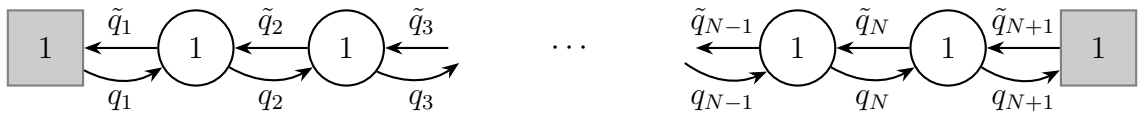
\begin{figure}[h!]
 \centering
\begin{tikzpicture} [place/.style={circle,draw=black!500,fill=white!20,thick,
inner sep=0pt,minimum size=10mm}, transition/.style={rectangle,draw=black!50,fill=black!20,thick,
inner sep=0pt,minimum size=10mm}, decoration={ markings,
mark=between positions 0.35 and 0.85 step 2mm with {\arrow{stealth}}}]
\node at ( -7,0) [transition] (0) {$1$};
\node at ( 7,0) [transition] (8) {$1$};
 \node at ( -5,0) [place] (1) {$1$};
 \node at (-4, -0.7) {$q_{2}$};\node at (-4, 0.35) {$\tilde{q}_{2}$};
 \node at ( 5,0) [place] (4) {$1$}; 
 
 \node at (4, -0.7) {$q_{N}$};\node at (4, 0.35) {$\tilde{q}_{N}$};
  \node at ( 3,0) [place] (3) {$1$}; 
  \node at (6, -0.7) {$q_{N+1}$};
  \node at (6, 0.35) {$\tilde{q}_{N+1}$};
  \node at (- 3,0) [place] (2) {$1$};
 \node at (-6, -0.7) {$q_{1}$};
 \node at (-6, 0.35) {$\tilde{q}_{1}$};
   \node at (-1.5,0) (5) {$ $}; \node at (-2, -0.7) {$q_{3}$};\node at (-2, 0.35) {$\tilde{q}_{3}$};
    \node at (1.5,0) (6) {$ $}; \node at (2, -0.7) {$q_{N-1}$};\node at (2, 0.35) {$\tilde{q}_{N-1}$};
     \node at (0,0) (7) {$\cdots$}; 
 \draw[->, -Stealth, thick] (0)[bend right] to (1);  
 \draw[->,-Stealth,thick] (1)--(0);
 \draw[->, -Stealth, thick] (4)[bend right] to (8);  
 \draw[->,-Stealth,thick] (8)--(4);
 \draw[thick, -Stealth] (4)--(3);\draw[thick, -Stealth] (3)[bend right] to(4);
 
 \draw[thick,Stealth-] (1)--(2);\draw[thick, Stealth-] (2)[bend left] to(1);
 
 \draw[thick, Stealth-] (2)--(5);\draw[thick, Stealth-] (-1.5, -0.2)[bend left] to(2);
 
 \draw[thick, -Stealth] (3)--(6);\draw[thick, -Stealth] (1.5, -0.2)[bend right] to(3);
\end{tikzpicture}
\caption{Theory B. For each node $i$, $i=1,...,N$, there is a $\mathcal{N}=2$ $U(1)$ vector multiplet $V_i$. Square nodes denote flavor symmetries. Oriented lines between adjacent nodes represent bifundamental chiral multiplets.}\label{Fig:theoryB}
\end{figure}

\paragraph{Theory B:}
Its mirror dual, which we name Theory B, is a linear quiver gauge theory (see Figure \ref{Fig:theoryB}) with $U(1)^N$ gauge group, $N+1$ bifundamentals $q_i,\tilde{q}_i$, $N+1$ singlets $s_i$ and superpotential
\begin{equation}\label{Eq:SupLinQuiv}
    W_B = \sum_{i=1}^{N+1} s_i q_i \tilde{q}_i \:.
\end{equation}
On this side, we can consider adding Polonyi terms, of the form
\begin{equation}
\delta W_{{\rm diag}} = \tilde \Phi_i^i \, s_i\,.
\end{equation}
where $\tilde \Phi$ is a background diagonal matrix. Equally importantly, as we will explain later, we can add monopole terms of the form
\begin{equation}\label{eq:mondef}
\delta W_{{\rm mon}} = \tilde \Phi_{\alpha^\star} \, w_{-\alpha}\,,
\end{equation} 
where the notation will be explained shortly.

Theory B has an $SU(N+1)$ acting on the Coulomb Branch coordinates. In particular, Theory B has a manifest topological $U(1)_T^N$ symmetry, which is expected to enhance to a $SU(N+1)$ global symmetry in the IR.\footnote{The arguments for this are solid for $\mathcal{N}=4$. However, the $\mathcal{N}=2$ case requires more indirect checks, such as identifying $SU(N+1)$ characters in the superconformal index \cite{Benvenuti:2024seb, Benvenuti:2025huk}.} The moment map for Theory B is given by $\mu$ in the adjoint of $SU(N+1)$, with $\mu_{ii}=s_i$, and the off-diagonal components are the monopole operators with charges in one-to-one correspondence with the roots of the $A_N$ Lie algebra associated with the topological symmetry.

Let us now fix our conventions for monopole operators. A monopole operator is specified by an $N$-tuplet of magnetic charges
\[\vec{m} = (m_1, \ldots, m_N) \in \mathbb{Z}^N\,.\]
Now, we recall the results of \cite{Bashkirov:2010kz, Gaiotto:2008ak}, that conclude that the IR R-charge of a monopole operator $w_{\vec{m}}$ with magnetic charges $\vec{m}$ will be given by\footnote{This applies to a balanced Abelian quiver.}
\begin{equation}
    \Delta(w_{\vec{m}}) = \frac{1}{2}\,\sum_{i=1}^N |m_i-m_{i+1}|\,.
\end{equation}
The monopole operators of interest, meaning those that lead to an enhanced global symmetry, are those with $\Delta=1$. The only possible solutions are monopoles with a single cluster of consecutive ones, regardless of the length. So, the simple `length-$1$' solutions are the following:
\begin{align}
    \vec{m}^{(1)}_1 = (1,0,0,\ldots,0)\\
    \vec{m}^{(1)}_2 = (0,1,0,\ldots,0)\\
    \ldots \\
    \vec{m}^{(1)}_N = (0,\ldots,0,1)\,,
\end{align}
These monopoles correspond to the simple roots of the enhanced flavor algebra. For `length-$2$' we have
\begin{align}
    \vec{m}^{(2)}_1 = (1,1,0,\ldots,0)\\
    \vec{m}^{(2)}_2 = (0,1,1,\ldots,0)\\
    \ldots \\
    \vec{m}^{(2)}_N = (0,\ldots,1, 1)\,,
\end{align}
and so on up to length-$(N-2)$, filling out the rest of the root system of the enhanced $SU(N+1)$.

To clarify the link between this basis of $U(1)$-charges, and the usual Dynkin labels for the adjoint representation of Lie$(G)$, we define the following basis $\mathfrak{h}_G = \langle \alpha_i^\ast\rangle$ for the Cartan subalgebra\footnote{These $\alpha^\star$ are often referred to as \emph{fundamental coweights}, not to be confused with the coroots.} $\mathfrak{h}_G$ such that
\begin{equation} \label{eq:fundcoweights}
    \alpha_i(\alpha_j^\ast) = \delta_{i j}\,.
\end{equation}
This is the appropriate language to describe monopole operators charged w.r.t. a single gauge node.
Given this, we will refer to a monopole $\omega_\alpha$ as the one charged under $\alpha^*$ only.

\paragraph{Mirror map} The mirror map works as follows:
\begin{equation}
   M_i^j = Q_i \tilde{Q}_j  \longleftrightarrow  \mu_{ij}  \,, \qquad V_+ \longleftrightarrow B\,,\qquad \qquad V_- \longleftrightarrow \tilde B \:, 
\end{equation}
where $V_\pm$ are the monopole operators of the $U(1)$ theory, while $B=q_1q_2...q_{N+1}$ and $\tilde B=\tilde q_1\tilde q_2...\tilde q_{N+1}$ are the baryons of the second theory. The diagonal $\mu_{ii} = s_i$, and the off-diagonal $\mu_{i j}$ are monopole operators pointing in the root space directions matching the meson of the A-side.

The deformations \eqref{eq:massdef} and \eqref{eq:mondef} are also mapped into each other via the following identifications:
\begin{align}
    \Phi &\leftrightarrow \tilde \Phi \\
    M_\alpha &\leftrightarrow w_\alpha\,,
\end{align}
where $M_\alpha$ is the A-side meson that points along the $\alpha$-root of $SU(N+1)$.

Let us now add to Theory B $N+1$ singlets $T_i$ and deform the superpotential \eqref{Eq:SupLinQuiv} to the following one
\begin{equation} \label{eq:bprime}
    W_{B'} = \sum_{i=1}^{N+1} s_i q_i \tilde{q}_i -  \sum_{i=1}^{N+1} s_i T_i \:.
\end{equation}
It is a mass term for $T_i$ and $s_i$. Integrating them out, one obtains a zero superpotential. The new theory will be called Theory B'.

We can obtain the mirror dual of Theory B', that we call Theory A', by adding the dual of the superpotential deformation to Theory A; on this side we then get the superpotential
\begin{equation}
    W = - \sum_{i=1}^{N+1}  T_i Q_i \tilde{Q}_i \:.
\end{equation}
Note, that $W_{B'}$ implies the chiral ring identification 
\begin{eqnarray}\label{eq:furtherid}
    \langle q_i \tilde q_i \rangle = \langle T_i \rangle\,.
\end{eqnarray}
 
\subsection{Monopole superpotentials}\label{Sec:MonopoleSuperpot}

In the following, we will analyze three-dimensional theories whose superpotentials include monopole operators. Since these operators are not functions of the fundamental fields, their effect on the IR effective theory requires special care. To address this, we will employ the techniques developed in \cite{Collinucci:2016hpz,Benini:2017dud,Collinucci:2017bwv}.

In this section, we focus on elementary examples that will serve as building blocks for deriving the effective superpotentials of the theories arising on D2-branes probing cDV singularities.

\subsubsection*{Building block I}
Let us consider a 3d $\mathcal{N}=2$ supersymmetric $U(1)$ gauge theory with two flavors and with the following superpotential that includes monopole operators:
\begin{equation}\label{Eq:SupN4LinQuivOneNode}
W = \varphi (q_1 \tilde{q}_1 - q_2 \tilde{q}_2) + w_- + P(\varphi) w_+ \:,
\end{equation}
where $P(\varphi)$ is a polynomial in $\varphi$.
This theory can be seen as a deformation of a 3d $\mathcal{N}=2$ abelian linear quiver gauge theory with one node and without superpotential. Let us think of it as taking the $B'$ theory in the language of \eqref{eq:bprime} plus deformation.

Its mirror can be found by taking the corresponding Theory A' and adding to its superpotential the dual of \eqref{Eq:SupN4LinQuivOneNode}. One obtains a $U(1)$ gauge theory with 2 flavors and superpotential
\begin{equation}\label{Eq:SupMirrorN4U12Fl}
    W = -T_1Q_1\tilde{Q}_1 - T_2Q_2\tilde{Q}_2 + \varphi (T_1-T_2) + P(\varphi)Q_1 \tilde{Q}_2 + Q_2\tilde{Q}_1  \:,
\end{equation}
where we have also used \eqref{eq:furtherid}.
The fields $Q_2$ and $\tilde{Q}_1$ are massive and can be integrated out: their equations of motion give 
\begin{equation}
    Q_2 = T_1 Q_1 \qquad \mbox{and} \qquad \tilde Q_1 = T_2 \tilde Q_2 \:.
\end{equation}
Plugging this into \eqref{Eq:SupMirrorN4U12Fl}, one obtains the effective superpotential\footnote{We do not integrate out the massive fields $T_1-T_2$ and $\varphi$. The reason for this choice will become clear when this building block is applied in the following sections.}
\begin{equation}
    W_{\rm eff} = (P(\varphi) - T_1T_2)Q_1\tilde Q_2 + \varphi(T_1-T_2) \:.
\end{equation}
One obtains a $U(1)$ gauge theory with one flavor. Its mirror dual is an XYZ model, with modified superpotential\footnote{In some of the cases analyzed in the following sections, the sign in front of $q_2,\tilde{q}_2$ in \eqref{Eq:SupN4LinQuivOneNode} will be $+$ instead of $-$. In such instances, it is sufficient to replace $T_2$ with $-T_2$.}
\begin{equation}
    W_{\rm eff} = X(Y\,Z + P(\varphi) - T_1T_2) + \varphi(T_1-T_2) \:.
\end{equation}

\subsubsection*{Building block II}
Let us now consider the more generic case, i.e. a 3d abelian linear quiver with $N$ nodes and with the following superpotential that includes monopole operators:
\begin{equation}\label{Eq:SupN4LinQuivNNode}
W = \sum_{i=1}^N \varphi_i (q_i \tilde{q}_i - q_{i+1} \tilde{q}_{i+1}) + \sum_{i=1}^N w_{-\alpha_i} + \left(\frac{1}{N}\sum_{i=1}^N \varphi_i \right) w_{+\alpha_{\rm high}} \:.
\end{equation}
where, as explained previously, $\omega_{-\alpha_i}$ refers to the monopole operator with charge $-1$ w.r.t. to the $\alpha_i$ node, and zero for the others. This theory can be seen as a deformation of a 3d $\mathcal{N}=2$ abelian linear quiver gauge theory with $N$ nodes and without superpotential. 

Its mirror can be found by taking the corresponding Theory A' and adding to its superpotential the dual of \eqref{Eq:SupN4LinQuivNNode}. One obtains a $U(1)$ gauge theory with $N+1$ flavors and superpotential
\begin{equation}\label{Eq:SupMirrorN4U1NFl}
    W = - \sum_{i=1}^{N+1} T_iQ_i\tilde{Q}_i + \sum_{i=1}^N \varphi_i (T_i - T_{i+1}) + \sum_{i=1}^N Q_{i+1}\tilde{Q}_i + \left(\frac{1}{N}\sum_{i=1}^N \varphi_i \right) Q_1\tilde{Q}_{N+1} \:.
\end{equation}
Integrating out the massive fields $Q_{i+1},\tilde Q_i$ with $i=1,...,N$ one obtains the effective superpotential
\begin{equation}\label{Eq:EffSupN4U1NFl}
    W_{\rm eff} = \left(\frac{1}{N}\sum_{i=1}^N \varphi_i - T_1T_2...T_{N+1}\right)Q_1\tilde Q_{N+1} + \sum_{i=1}^N \varphi_i (T_i - T_{i+1}) \:.
\end{equation}
Hence, we have a $U(1)$ gauge theory with one flavor and superpotential \eqref{Eq:EffSupN4U1NFl}. Its mirror dual is an XYZ model, with modified superpotential
\begin{equation}
    W_{\rm eff} = X\,\left(Y\,Z + \frac{1}{N}\sum_{i=1}^N \varphi_i  -  T_1T_2...T_{N+1}\right) + \sum_{i=1}^N \varphi_i (T_i - T_{i+1}) \:.
\end{equation}

\section{cDV singularities and the Higgs field}\label{Sec:cDVHiggs}

\subsection{Summary of cDV}
In this paper we consider a class of Calabi-Yau (CY) threefolds, the compound Du Val (cDV) singularities.
They are hypersurfaces in $\mathbb{C}^4$ of the form
\begin{equation}\label{Eq:cDVsing}
   P_{\mathcal G}(x,y,z) + w \, g(x,y,z,w) =0\:,
\end{equation}
with $P_{\mathcal G}(y,z)$ following the ADE classification:
\begin{equation}\label{ADE singularities}
\begin{split}
& P_{A_n}=x^2+y^2+z^{n+1}\,,\\
& P_{D_n}=x^2+z y^2+z^{n-1}\,,\\
& P_{E_6}=x^2+y^3+z^4 \,,\\
& P_{E_7}=x^2+y^3+yz^3 \,,\\
& P_{E_8}=x^2+y^3+z^5 \,.\\
\end{split}
\end{equation}
The first three monomials in \eqref{Eq:cDVsing} reconstruct the ADE singularity of type $\mathcal{G}$.
ADE singularities $P_{\mathcal{G}}(x,y,z)=0$ are surface singularities whose resolutions replace the singular point with a collection of holomorphic two-spheres intersecting according to the Dynkin diagram of the associated ADE Lie algebra. With a slight abuse of notation, we will refer to these exceptional spheres as the simple roots. Such surface singularities can also be deformed, by adding further monomials to the surface equations. 

The last term in \eqref{Eq:cDVsing} can actually be interpreted as a deformation of this singularity, in which the coefficient of the added monomials depend on an additional complex coordinate $w$. 
This gives rise to a one-parameter family of deformed ${\mathcal G}$-singularities, fibered over a complex plane $\mathbb{C}_w$: the fiber at $w=0$ develops an ADE singularity, while at $w\neq 0$ the ADE singularity is deformed. The singular point of the ALE fiber, over $w = 0$, corresponds to the singular point of the threefold.

Depending on how the coefficients of the deformed ADE singularity vary with $w$, the resolution of the threefold singularity to a smooth space may blow up either all or only a subset of the simple roots. This phenomenon is referred to as a (partial) simultaneous resolution. The roots that remain unresolved are conventionally indicated (or marked) in the Dynkin diagram associated with the ADE algebra. For further details on these threefolds, see \cite{katz1992gorensteinthreefoldsingularitiessmall}.

\subsection{Higgs field} \label{subsec:higgsfield}
In a series of papers \cite{Collinucci:2021wty,Collinucci:2021ofd,Collinucci:2022rii,DeMarco:2022dgh}, these singularities have been described in terms of a Higgs field $\Phi(w)$.\footnote{See \cite{Sangiovanni:2024nfz} for generalizations to four-folds.} This is a scalar field in the adjoint representation of the ADE algebra, depending holomorphically on the complex coordinate $w$. Given such a field, one can reconstruct the hypersurface equation, understand which roots are simultaneously resolved and compute the corresponding Gopakumar-Vafa invariants.

For the A- and D-type cDV threefolds that we consider in this paper, the hypersurface equations induced by $\Phi(w)$ are
\begin{equation}\label{Eq:ADequationsFromPhi}
\begin{split}
& A_n: \qquad  x^2+y^2 = \det(z\mathbb{1}_{n+1}-\Phi)\\
& D_n: \qquad  x^2 + zy^2 -\frac{\sqrt{\det(z\mathbb{1}_{2n}+\Phi^2)}-\mbox{Pfaff}^2(\Phi)}{z} +2y\mbox{Pfaff}(\Phi)=0
\end{split}
\end{equation}
where $\Phi$ is written as a matrix in the fundamental representation of $A_n$ or $D_n$ respectively.

As  $w$ varies, the Higgs field $\Phi(w)$ takes values in a subalgebra $\mathcal{L}$  of the ADE Lie algebra (a detailed discussion of this construction can be found in \cite{Collinucci:2022rii}, which we briefly summarize in the following). The commutant $\mathcal{H}$ of $ \mathcal{L}$ — that is, the set of elements commuting with all of $\mathcal{L}$ — is an abelian subalgebra contained in the Cartan subalgebra $\mathcal{C}$.\footnote{The reason for this is traced back to the M-theory origin of the field $\Phi$ and it is explained in \cite{Collinucci:2021ofd,Collinucci:2022rii}.} By evaluating the roots $\alpha$ on elements $h \in \mathcal{H}$, one can determine which simple roots are resolved: if $\alpha(h) \neq 0$ for some $h \in \mathcal{H}\subset \mathcal{C}$, then the root $\alpha$ is a vanishing two-cycle that can be resolved in the (partial) simultaneous resolution.  
The subalgebra $\mathcal{L}$ is known as a \emph{Levi subalgebra}, since it commutes with an abelian subalgebra.
This implies that the resolution of the threefold induced by $\Phi$ can be understood by identifying the Levi subalgebra to which it belongs.

The Levi subalgebra has the following form
\begin{equation}
    \mathcal{L}=\bigoplus_\lambda \mathcal{L}_\lambda \oplus \mathcal{H}
\end{equation}
where $\mathcal{L}_\lambda$ are simple algebras and $\mathcal{H}$ is the abelian commutant of $\mathcal{L}$. We will choose the following basis for $\mathcal{H}$: take the basis $\alpha_1^\ast,...,\alpha_r^\ast$ for the Cartan subalgebra of the ADE Lie algebra of rank $r$, where $\alpha_i^\ast$ are defined by $\alpha_i(\alpha_j^\ast)=\delta_{ij}$ ($\alpha_i$ $i=1,...,r$ are the simple roots); then $\mathcal{H}=\langle \alpha_1^\ast,...,\alpha_k^\ast \rangle$ with $\alpha_1,...,\alpha_k$ ($k\leq r$) the simple roots that are blown up in the (partial) simultaneous resolution of the threefold.
One can define invariant coordinates on the Levi subalgebra as follows: for each simple summand, choose the Casimir invariants of that algebra;\footnote{For an $A_n$ summand, an element $g \in A_n$ has invariant coordinates $\varrho_{\lambda_q} = \mathrm{Tr}\, g^q$.} the remaining coordinates are given by the coefficients in the expansion of $\Phi$ along the basis $(\alpha_1^\ast, \dots, \alpha_k^\ast)$ of $\mathcal{H}$.
For example, consider $\mathcal{L} = A_1^{(1)} \oplus A_1^{(2)} \oplus \langle \alpha_3^\ast \rangle$. For a generic element $\Phi \in \mathcal{L}$, we have: (i) invariant coordinate $\varrho_1 = \tfrac{1}{2}\tr(\Phi_1^2)$ for the first $A_1$ summand, (ii) invariant coordinate $\varrho_2 = \tfrac{1}{2}\tr(\Phi_2^2)$ for the second $A_1$ summand, and (iii) invariant coordinate $\varrho_3$ (coefficient along $\alpha_3^\ast$) for the $U(1)$ part. A generic $\Phi \in \mathcal{L}$ can then be written as $\Phi(\varrho_1,\varrho_2,\varrho_3)$, and specifying a base change $\varrho_i = f_i(w)$ gives a specific cDV threefold.

As explained in \cite{Collinucci:2022rii,Moleti:2024skd}, a generic element $\Phi \in \mathcal{L}$ can be expressed in terms of the invariant coordinates, i.e., $\Phi(\varrho)$. One can then substitute this $\Phi$ into \eqref{Eq:ADequationsFromPhi} (for the A- and D-type cases; analogous formulae exist for E-type cases) to obtain a family of deformed ADE surfaces with a simultaneous resolution pattern determined by the Levi subalgebra to which $\Phi$ belongs. This family is fibered over the space parametrized by the invariants $\varrho$ of the Levi subalgebra $\mathcal{L}$. Specifying a {\it base change} $\varrho_a = f_a(w)$ then yields a threefold with the same resolution pattern as that of the original family.

Hence, a $\Phi(w)$, and correspondingly a cDV threefold, is given by the choice of a colored ADE Dynkin diagram (the resolution pattern), or equivalently a Levi subalgebra, and a base change $\varrho_a=f_a(w)$. 


In summary, a cDV threefold can be viewed as an ADE surface fibration over $\mathbb{C}_w$, where each simple root $\alpha_i$ corresponds to a vanishing 2-sphere in the fiber over $w=0$. As we move around a closed loop encircling $w=0$ in the base, there are two possible behaviors for these spheres. 
The sphere associated with $\alpha_i$ may return to itself after traversing the loop. In this case, it extends globally as a holomorphic cycle in the threefold and can be resolved. 
Alternatively, the sphere may undergo monodromy, mixing non-trivially with other spheres as we encircle $w=0$. In this case, it does not lift to a globally well-defined holomorphic curve in the threefold. 
Geometries in which some spheres exhibit the latter behavior are referred to as \emph{monodromic} cDV threefolds, while those in which all spheres behave as in the former case are called \emph{non-monodromic} \cite{Cachazo:2001aa,Cachazo:2001gh}.

The resolution pattern is thus encoded in a colored Dynkin diagram: white nodes correspond to resolvable roots (globally defined spheres), while colored nodes correspond to non-resolvable roots (spheres undergoing monodromy). Connected subdiagrams of colored nodes correspond to simple summands of the Levi subalgebra $\mathcal{L}$.

Algebraically, this structure is encoded in the Higgs field $\Phi(w)$: a root $\alpha$ is colored if and only if $\Phi$ contains off-diagonal components—the step operators $e_{\pm\alpha}$—beyond just Cartan elements. For example, $\Phi(w) = e_\alpha + \varrho(w) e_{-\alpha}$ contains step operators, indicating that $\alpha$ is colored and undergoes monodromy. In contrast, $\Phi(w) = \varrho(w) \alpha^\ast$ lies purely in the Cartan, indicating that $\alpha$ is white (resolvable).

\section{D2-brane probing a cDV threefold singularity}\label{Sec:D2onADEfib}

The central claim of this paper is that a broad class of cDV threefold singularities can be reconstructed as moduli spaces of suitably deformed three-dimensional probe theories. More precisely, we start from the $\mathcal N=4$ worldvolume theory of a D2-brane probing the corresponding undeformed ADE surface singularity, and we deform it by couplings determined by the Higgs field $\Phi(w)$ that encodes the threefold geometry.

In non-monodromic cases, where $\Phi(w)$ lies in the Cartan subalgebra, the deformation is expressed directly in terms of polynomial superpotential couplings. In monodromic cases, the off-diagonal components of $\Phi(w)$ generate monopole superpotential terms associated with the corresponding colored Dynkin subdiagrams. The technical task is then to convert these monopole deformations into an effective description with ordinary polynomial interactions by means of local 3d mirror symmetry and ungauging/regauging \cite{Collinucci:2014qfa,Collinucci:2016hpz}.

The outcome is an effective three-dimensional $\mathcal N=2$ theory whose Higgs branch\footnote{Throughout the paper, when referring to the Higgs branch of the 3d $\mathcal{N}=2$ theory, we implicitly mean the branch that matches the moduli space of the corresponding 4d worldvolume theory on a D3-brane. As discussed in Section~\ref{Sec:D3vsD2}, the D2-brane description arises from compactifying the D3-brane on a circle; the additional 3d Coulomb-branch directions associated with the circle are frozen when identifying the geometric moduli space.} reproduces the cDV threefold in the examples studied below. Upon restricting to the origin of the Coulomb branch and taking the decompactification limit discussed in Section~\ref{Sec:D3vsD2}, this same branch is the one relevant to the four-dimensional D3-brane probe theory.

In order to derive the 3d $\mathcal{N}=2$ worldvolume theory on D2-branes probing cDV threefold singularities from the geometric data encoded in the Higgs field $\Phi(w)$, we adopt the following three-step strategy:
\begin{enumerate}
    \item Start with the 3d $\mathcal{N}=4$ quiver for a D2-brane probing the ADE surface 
   singularity.
    \item Deform the superpotential using $\Phi(w)$.
    \item Integrate out massive fields to obtain the effective IR theory. The classical moduli space of this 3d theory reproduces the probed cDV threefold. Upon decompactification to 4d, this theory flows to the worldvolume theory of a D3-brane probing the same singularity.
\end{enumerate}
For non-monodromic fibrations (all roots resolved), $\Phi$ belongs to the Cartan subalgebra and induces a simple deformation. For monodromic cases (colored nodes), $\Phi$ contains nilpotent 
elements that induce monopole operators in the superpotential.

Our analysis is formulated most concretely for cDV threefolds that admit an A- or D-type description in terms of the Higgs field $\Phi(w)$ and for which the associated probe theory can be treated in terms of abelian quiver building blocks. In these cases the map from $\Phi(w)$ to a deformation of the D2-brane theory can be implemented explicitly.

As we will discuss in a moment, in monodromic cases our treatment relies on local 3d mirror symmetry and on isolating suitable balanced abelian subquivers. The resulting effective description is therefore controlled in the class of examples in which this procedure can be implemented explicitly. Our examples provide strong evidence that this prescription captures the correct geometry, but we do not claim here a uniform derivation for arbitrary non-abelian Levi blocks.


\subsection{D2-brane probing an ADE singularity}\label{Sec:D2onADE}

We begin by reviewing the worldvolume theory on a D2-brane probing an ADE singularity.
It is a $\mathcal{N}=4$ 3d supersymmetric quiver gauge theory.

\begin{figure}[H]
 \centering
\begin{tikzpicture} [place/.style={circle,draw=black!500,fill=white!20,thick,
inner sep=0pt,minimum size=10mm}, transition/.style={rectangle,draw=black!50,fill=black!20,thick,
inner sep=0pt,minimum size=10mm}, decoration={ markings,
mark=between positions 0.35 and 0.85 step 2mm with {\arrow{stealth}}}]
 \node at ( -5,0) [place] (1) {$1$};\node at (-5, -0.8) {$\varphi_1$};\node at (-4, -0.7) {$q_{2}$};\node at (-4, 0.35) {$\tilde{q}_{2}$};
 \node at ( 5,0) [place] (4) {$1$}; \node at (5, -0.8) {$\varphi_{r}$};\node at (4, -0.7) {$q_{r}$};\node at (4, 0.35) {$\tilde{q}_{r}$};
  \node at ( 3,0) [place] (3) {$1$}; \node at (3, -0.8) {$\varphi_{r-1}$};\node at (3, 3.4) {$q_{r+1}$};\node at (3, 2.2) {$\tilde{q}_{r+1}$};
  \node at (- 3,0) [place] (2) {$1$}; \node at (-3, -0.8) {$\varphi_2$};
  \node at (0,4)[place] (0) {$1$}; \node at (0, 4.8) {$\varphi_{r+1}$};\node at (-3, 3.3) {$q_{1}$};\node at (-3, 2) {$\tilde{q}_{1}$};
   \node at (-1.5,0) (5) {$ $}; \node at (-2, -0.7) {$q_{3}$};\node at (-2, 0.35) {$\tilde{q}_{3}$};
    \node at (1.5,0) (6) {$ $}; \node at (2, -0.7) {$q_{r-1}$};\node at (2, 0.35) {$\tilde{q}_{r-1}$};
     \node at (0,0) (7) {$\cdots$}; 
 \draw[->, -Stealth, thick] (0)[bend right] to (1);  \draw[->,-Stealth,thick] (1)--(0);\draw[->,-Stealth,thick] (4)[bend right] to (0);\draw[->,-Stealth,thick] (0)--(4);
 \draw[thick, -Stealth] (4)--(3);\draw[thick, -Stealth] (3)[bend right] to(4);
 
 \draw[thick,Stealth-] (1)--(2);\draw[thick, Stealth-] (2)[bend left] to(1);
 
 \draw[thick, Stealth-] (2)--(5);\draw[thick, Stealth-] (-1.5, -0.2)[bend left] to(2);
 
 \draw[thick, -Stealth] (3)--(6);\draw[thick, -Stealth] (1.5, -0.2)[bend right] to(3);
\end{tikzpicture}
\caption{$A_r$ theory. For each node $i$, $i=1,...,r+1$, there is a $\mathcal{N}=4$ $U(1)$ vector multiplet $V_i$ containing a $\mathcal{N}=2$ vector multiplet and an adjoint chiral $\varphi_i$. Pairs of oriented lines between adjacent nodes represent bifundamental hypermultiplets $(q_i, \tilde{q}_i)$.}\label{Fig:ArQuiver}
\end{figure}
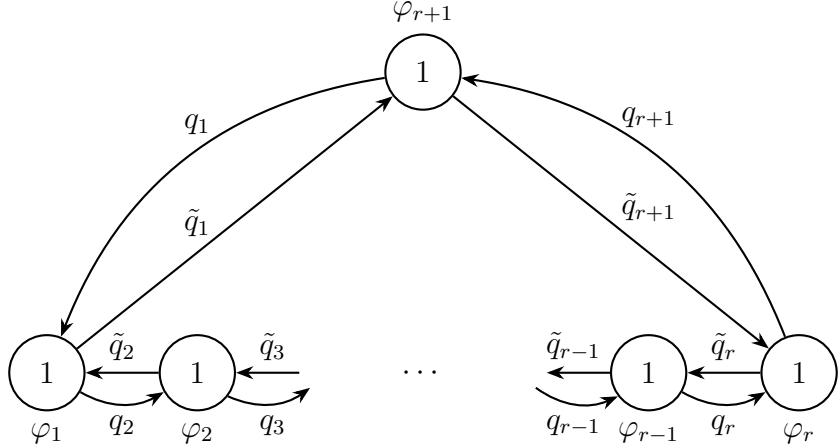

The quiver has the shape of the affine Dynkin diagram of the corresponding ADE Lie algebra \cite{Douglas:1996sw}. 
The nodes of the quiver correspond to fractional D2-branes; the gauge group at each node is $U(n_i)$, where $n_i$ is the dual Coxeter label of the $i$-th node in the associated Dynkin diagram.
Each pair of arrows connecting the nodes in the quiver is a bifundamental hypermultiplet. As an example, Figure~\ref{Fig:ArQuiver} illustrates the $A_r$ quiver.  
In this case, the gauge group is~$\left(\prod_{i=0}^r U(1)\right)/U(1)$. The vector multiplet chiral fields $\phi_i$ are coupled to the hypermultiplets  $(q_i,\tilde{q}_i)$ via the $\mathcal{N}=4$ superpotential
\begin{equation}\label{eq:N4sup}
W=\sum_{i=1}^{r+1} (\varphi_i-\varphi_{i-1})q_i\tilde{q}_i, \:,
\end{equation}
with $\phi_{r+2}\equiv \phi_1$. There is also a D-term potential that produces the D-term relations
\begin{equation}\label{eq:D-termAr}
 |q_i|^2+ |\tilde{q}_{i-1}|^2 - |\tilde{q}_{i}|^2 - |q_{i-1}|^2= 0 \qquad\qquad i=1,...,r \:.
\end{equation}

The $A_r$ theory features $r$ triplets of Fayet–Iliopoulos (FI) parameters: the triplet can be separated
in a complex FI-parameter that appears in the superpotential, and a real one that modifies the relations \eqref{eq:D-termAr}; these can be regarded as mass parameters (i.e. background vector multiplets) for the $U(1)_T^r$ topological symmetry. 

As we said above, for $\mathcal{N}=4$ quiver gauge theories, monopole operators and the scalar fields $\varphi_i$ can be assembled into the moment map $\mu$ for the topological symmetry. $\mu$ is an object transforming in the adjoint representation of the Lie algebra $\mathcal{G}$. 
Under mirror symmetry, the $A_r$ quiver is mapped to a $U(1)$ gauge theory with $r+1$ flavors that has a manifest $SU(r+1)$ symmetry, with moment map $\mu^{\rm mirror}_{ij} = Q_i\tilde Q_j$. As we discussed above, mirror symmetry maps the moment map $\mu$ of the original theory to the moment map $\mu^{\rm mirror}$ in the dual theory.

\subsection{Non-monodromic ADE fibrations}\label{Sec:D2onNonmonodFib}

The theory of a D2-brane probing the threefold singularity $X$ can be viewed as a deformation of the theory describing a D2-brane probing the corresponding ADE surface singularity. This deformation is governed by the \emph{background} Higgs field~$\Phi(w)$, which is non-dynamical from our 3d point of view and should not be confused with the dynamical adjoint scalars $\varphi_i$ in the quiver gauge theory.

\subsubsection*{Fractional branes and holomorphic volumes}

To understand the superpotential deformation, we first recall the classic result of \cite{Cachazo:2001gh} (building on earlier work \cite{Witten:1997ep,Kachru:2000ih,Aganagic:2000gs}). Consider D5-branes wrapping 2-cycles in a fibered ADE geometry over a base $\mathbb{C}_w$. In the absence of background fields, the D5-branes give rise to an $\mathcal{N}=2$ gauge theory with FI parameters $\alpha_i$ equal to the holomorphic volumes of the wrapped 2-cycles. When a non-trivial fibration is introduced, a superpotential is generated in the worldvolume theory.

The key observation of \cite{Cachazo:2001gh} is that the BPS tension of domain walls, computed by integrating the holomorphic three-form $\Omega = \omega \wedge dw$ over a three-chain, determines the superpotential:
\begin{equation}\label{Eq:CVformula}
    W(\varphi) = \int_{w=\Phi} \alpha(w)\,dw\:,
\end{equation}
where $\varphi$ is the scalar field parametrizing the position of the fractional brane along the base, and $\alpha(w) = \int_{S^2(w)} \omega$ is the holomorphic volume of the wrapped cycle at position $w$. Here $\omega$ denotes the holomorphic $(2,0)$-form of the ADE fiber. This formula implies that
\begin{equation}
    \frac{\partial W}{\partial \varphi} = \alpha(\varphi)\:,
\end{equation}
i.e., the derivative of the superpotential reproduces the holomorphic volume of the wrapped cycle at the brane's position.

Conversely, if one wishes to engineer a gauge theory with a specified superpotential $W(\varphi) = P(\varphi)$, the fibration must be chosen such that $\alpha(w) = P'(w)$. This establishes a direct dictionary between the geometric deformation of the ADE fiber and the superpotential of the gauge theory.

In our setup, fractional D4-branes (arising from the D2-brane probe splitting into components) wrap the vanishing spheres associated with each simple root $\alpha_i$ of the ADE Dynkin diagram. The Cachazo-Vafa formula \eqref{Eq:CVformula} will guide our construction of the superpotential deformation induced by the background Higgs field $\Phi(w)$.

\subsubsection*{Deriving the deformation from $\Phi(w)$}

We now formulate the basic proposal that translates the geometric data encoded in $\Phi(w)$ into a deformation of the D2-brane worldvolume superpotential. The proposal is motivated by the holomorphic-volume formula \eqref{Eq:CVformula}, by the interpretation of the fractional branes as branes wrapping the vanishing two-cycles of the ADE fiber, and by consistency with the mirror-symmetry treatment of the monodromic examples discussed later on. In the examples analyzed in this paper, this prescription reproduces the expected cDV geometry.

For balanced quivers (where each node has rank $N_i$ and $2N_i$ flavors), consider isolating a single fractional brane at node $i$ together with its $2N_i$ hypermultiplets. This subsystem supports a $U(N_i)$ gauge group with $\mathcal{N}=4$ supersymmetry. The Coulomb branch of this subsystem possesses an $SU(2)$ flavor symmetry—distinct from the $SU(2N_i)$ flavor symmetry acting on the Higgs branch—which corresponds to the $\mathfrak{su}(2)_{\alpha_i}$ subalgebra of $\mathcal{G}$ generated by 
$$
\langle e_{\alpha_i},e_{-\alpha_i},h_{\alpha_i}\equiv [e_{\alpha_i},e_{-\alpha_i}]\rangle.
$$

The background Higgs field $\Phi$, which transforms in the adjoint representation of $\mathcal{G}$, couples to the fractional brane worldvolume theory through the moment map $\mu_{\alpha_i}$ associated to this Coulomb branch flavor symmetry $\mathfrak{su}(2)_{\alpha_i}$. In the non-monodromic case, where $\Phi$ lies entirely in the Cartan subalgebra, this coupling involves only the Cartan part of $\mu_{\alpha_i}$, namely $\varphi_{\alpha_i} h_{\alpha_i}$, where $\varphi_{\alpha_i}$ is the scalar field parametrizing the position of the fractional brane along the base.

The Higgs field admits a polynomial expansion
\begin{equation}\label{Eq:SupDefNonMonod}
    \Phi(w) = \sum_{\ell=0}^k \Phi_\ell \, w^\ell\:,
\end{equation}
where the brane's position in the base is $w = \varphi_i$. The coupling between $\Phi(\varphi_i)$ and the moment map generates an induced superpotential in the worldvolume theory:
\begin{equation}
    \delta W_i =  \sum_{\ell=0}^k \frac{\varphi_i^{\ell}}{\ell+1} \, \kappa \left( \Phi_\ell \,,\,  \mu_{\alpha_i}   \right)
    = \sum_{\ell=0}^k \frac{\varphi_i^{\ell+1}}{\ell+1} \, \kappa \left( \Phi_\ell \,,\,  h_{\alpha_i}   \right) \:,
\end{equation}
where $\kappa$ denotes the Killing form. The factor $(\ell+1)^{-1}$ is chosen precisely to satisfy the Cachazo-Vafa condition \eqref{Eq:CVformula}:
\begin{equation}\label{Eq:DerivDeltaW_CV}
    \frac{\partial \delta W_i}{\partial \varphi_i} = \sum_{\ell=0}^k \varphi_i^\ell \, \kappa(\Phi_\ell, h_{\alpha_i}) = \kappa\left(\Phi(\varphi_i), h_{\alpha_i}\right) = \alpha_i\left(\Phi(\varphi_i)\right)\:,
\end{equation}
where in the last step we have used the normalization of the Killing form for ADE algebras, $\kappa(h_\alpha, h_\beta) = \alpha \cdot \beta$. 

This has a clear geometric interpretation: at a given value $w$ in the base, the holomorphic volume of the sphere $\alpha_i$ in the deformed fiber is $\alpha_i(\Phi(w))$. Since the fractional brane wrapping $\alpha_i$ sits at position $w = \varphi_i$, equation \eqref{Eq:DerivDeltaW_CV} correctly reproduces the holomorphic volume, in agreement with \cite{Cachazo:2001gh}.

\ 


In the following, we examine some well-known examples and verify that the proposed deformation indeed reproduces the expected geometry in the moduli space.

\subsubsection*{Examples 1: Conifold}

We begin with the simplest and most familiar example, namely the conifold, given by the hypersurface
\begin{equation}\label{Eq:equationConifold}
    u v = z^2 - w^2 \:.
\end{equation}

This threefold can be regarded as a family of deformed $A_1$ singularities, where $w$ is the deformation parameter. The corresponding $\Phi(w)$ is the following $2\times 2$ matrix \cite{Collinucci:2021ofd}
\begin{equation}\label{Eq:PhiConifold}
    \Phi(w)=\begin{pmatrix}
        w & 0 \\ 0 & -w
    \end{pmatrix} = 2w \alpha^\ast \:.
\end{equation}
Plugging this into \eqref{Eq:ADequationsFromPhi} one reproduces correctly \eqref{Eq:equationConifold}. 

Let us probe the conifold singularity by a D2-brane. According to what we have said in this section, the resulting theory is the $A_1$ quiver gauge theory deformed by 
\begin{eqnarray}
    \delta W_1 &=& \frac{\varphi_1}{2} \alpha(\Phi(\varphi_1)) = \varphi_1^2 \,, \\
    \delta W_2 &=& - \frac{\varphi_2}{2} \alpha(\Phi(\varphi_2)) = -\varphi_2^2 \:.
\end{eqnarray}
Here we have used that the fractional brane corresponding to the extended node in the affine Dynkin diagram is wrapping $\alpha_2=-\alpha$.

Now, let us consider the complete deformed superpotential:
\begin{eqnarray}
 \begin{aligned}
    W &= (\varphi_1-\varphi_2)(q_1\tilde{q}_1-q_2\tilde{q}_2) + \varphi_1^2-\varphi_2^2 
    = \varphi_-(q_1\tilde{q}_1-q_2\tilde{q}_2+ \varphi_+)\:,
 \end{aligned}
\end{eqnarray}
where we have redefined $\varphi_-=\varphi_1-\varphi_2$ and $\varphi_+=\varphi_1+\varphi_2$.
Notice that this deformed superpotential for the conifold was already introduced by Klebanov-Witten \cite{Klebanov:1998hh}, as noticed in \cite{Cachazo:2001gh}.

The fields $\varphi_\pm$ are massive. Integrating them out, we get a quiver gauge theory, with the $A_1$ quiver and zero superpotential, i.e. we obtain exactly the theory of a D-brane probing the conifold.

We have verified that the moduli space reproduces the conifold geometry. 
This confirms our prescription \eqref{Eq:SupDefNonMonod} in the simplest case where $\Phi(w)$ contains only linear terms. The next examples will test the method 
for higher-degree base changes.

\subsubsection*{Example 2: Reid's Pagodas as a family of deformed $A_1$ surfaces}\label{Sec:ReidsPagodaNonMonodromicA1}

Let us now consider the following three-fold, known as Reid's pagoda:
\begin{equation}\label{Eq:equationReidsPagoda}
    u v = z^2 - w^{2k} \:.
\end{equation}
This is again a non-monodromic $A_1$-fibration (that reduces to the conifold for $k=1$), whose corresponding Higgs field is
\begin{equation}\label{Eq:PhiPagodaA1}
    \Phi(w)=\begin{pmatrix}
        w^k & 0 \\ 0 & -w^k
    \end{pmatrix} = 2w^k \alpha^\ast \:.
\end{equation}
This induces the superpotential deformations
\begin{eqnarray}
    \delta W_1 &=& \frac{\varphi_1}{k+1} \alpha(\Phi(\varphi_1)) = \frac{2}{k+1} \varphi_1^{k+1} \\
    \delta W_2 &=& - \frac{\varphi_2}{k+1} \alpha(\Phi(\varphi_2)) = -\frac{2}{k+1}\varphi_2^{k+1} \:.
\end{eqnarray}
We obtain, then an $A_1$ quiver with the superpotential
\begin{eqnarray}
 \begin{aligned}\label{Eq:PagodaNonMonodromicSuperpot}
    W &= (\varphi_1-\varphi_2)(q_1\tilde{q}_1-q_2\tilde{q}_2) + \frac{2}{k+1}\varphi_1^{k+1}-\frac{2}{k+1}\varphi_2^{k+1} \:.
 \end{aligned}
\end{eqnarray}
This superpotential appeared in \cite{Cachazo:2001gh,Cachazo:2001aa} and we will rederive it in the following in a different way, by considering the Reid's Pagoda as a family of deformed $A_{2k-1}$ singularities (by exchanging the roles of $z$ and $w$ in \eqref{Eq:equationReidsPagoda}).

Let us check that the F-term moduli space is the CY threefold \eqref{Eq:equationReidsPagoda}. The F-terms are
\begin{eqnarray}
    \begin{aligned}
        \partial_{\varphi_1}W=0&: \qquad q_1\tilde{q}_1-q_2\tilde{q}_2 + 2 \varphi_1^k =0 \\        \partial_{\varphi_2}W=0&: \qquad q_1\tilde{q}_1-q_2\tilde{q}_2 + 2 \varphi_2^k =0 \\
        \partial_{q_i}W=0&: \qquad (\varphi_1-\varphi_2)\,\tilde{q}_i=0 \qquad i=1,2\\
        \partial_{\tilde{q}_i}W=0&: \qquad (\varphi_1-\varphi_2)\,q_i=0 \qquad i=1,2 \\
    \end{aligned}
\end{eqnarray}
The solutions are given by $\varphi_2=\varphi_1\equiv\varphi$ and $q_2\tilde{q}_2=q_1\tilde{q}_1+2\varphi^k$. The moduli space is then given by defining gauge invariant combinations and finding the relations they need to satisfy. Here we have the following four gauge invariants (plus the field $\varphi$):
\begin{equation}
    \begin{aligned}
        U \equiv q_1q_2 ,\qquad V\equiv\tilde{q}_1\tilde{q}_2, \qquad
        X_1\equiv q_1\tilde{q}_1,\qquad 
        X_2\equiv q_2\tilde{q}_2
    \end{aligned}
\end{equation}
They satisfy the obvious relation $U\,V=X_1\,X_2$. Imposing the relation coming from the F-term, i.e. $X_2=X_1+2\varphi^k$, we have
\begin{eqnarray}
    U\,V=X_1\,(X_1+2\varphi^k) \:,
\end{eqnarray}
that can be brought in the form \eqref{Eq:equationReidsPagoda} by redefining the coordinates as $X_1=Z-\varphi^k$.

Notice that $\varphi$ plays the role of $w$. At a generic point of the moduli space $\varphi_1=\varphi_2=\varphi$, i.e. the fractional branes move (as a bound state) away from the origin at $w=\varphi$, that is consistent with the interpretation of the theory of D-branes probing an ADE surface singularity fibered over the complex plane $\mathbb{C}_w$.

This example demonstrates that our method correctly handles polynomial 
base changes $\varrho(w)=w^k$. Notice that the effective superpotential now 
contains terms $\varphi^{k+1}$, reflecting the degree of the base change. We 
will revisit this geometry in the next section from a different perspective 
(as a monodromic $A_{2k-1}$ fibration).

\subsubsection*{Examples 3: Generalized conifold}

The final example of non-monodromic fibration that we consider is the following hypersurface equation
\begin{eqnarray}\label{Eq:equationGenCon}
    uv=z(z-w)(z+w) \:.
\end{eqnarray}
This is a family of $A_2$ deformed singularities.

The corresponding Higgs field is
\begin{equation}\label{Eq:PhiGenConifold}
    \Phi(w)=\begin{pmatrix}
        w &  & \\ & -w & \\ & & 0 \\
    \end{pmatrix} = w ( 2\alpha_1^\ast - \alpha_2^\ast) \:.
\end{equation}
The induced superpotential deformations are
\begin{eqnarray}
    \delta W_1 &=& \frac{\varphi_1}{2} \alpha_1(\Phi(\varphi_1)) = \varphi_1^{2}\:, \\
    \delta W_2 &=& \frac{\varphi_2}{2} \alpha_2(\Phi(\varphi_2)) = -\frac{1}{2}\varphi_2^2\:,\\
    \delta W_3 &=& -\frac{\varphi_3}{2} (\alpha_1(\Phi(\varphi_3))+\alpha_2(\Phi(\varphi_3))) = -\frac{1}{2}\varphi_3^2 \:.
\end{eqnarray}
We then obtain the $A_2$ quiver with the following superpotential
\begin{equation}
W=\sum_{i=1}^{3} (\varphi_i-\varphi_{i-1})q_i\tilde{q}_i +\varphi_1^2 -\frac{1}{2}\varphi_2^2-\frac{1}{2}\varphi_3^2\:.
\end{equation}
This is the same superpotential found in \cite{Cachazo:2001gh}. 
By simple computations, one shows that the resulting moduli space reproduces the geometry \eqref{Eq:equationGenCon}, with $\varphi_1=\varphi_2=\varphi_3\equiv\varphi$ that correctly plays the role of $w$.

Unlike the previous $A_1$ cases, this example involves multiple roots, 
showing how different fractional branes couple independently to 
different components of $\Phi$. The next section addresses the more 
challenging monodromic cases where $\Phi$ contains off-diagonal terms.

\subsection{Monodromic ADE fibrations}\label{Sec:D2onMonodFibr}

In monodromic fibrations, some vanishing spheres of the ADE fiber do not extend to globally defined holomorphic curves in the total space of the threefold. Equivalently, as one goes around the singular point in the base, these spheres undergo non-trivial monodromy. 
This data is encoded in a \emph{colored Dynkin diagram}. We color the nodes of such curves, and leave white those that will be small-resolvable curves in the threefold. 

\begin{figure}[h]
\centering
\begin{tikzpicture}[
  node/.style={circle,draw,inner sep=0pt,minimum size=6pt},
  edge/.style={line width=0.6pt}
]

\begin{scope}[shift={(0,0)}]
  \node at (0,2.35) {$(a)$};

  \node[node, fill=black] (a1) at (0,0) {};
  \node[node] (a4) at (1,0) {};
  \node[node, fill=black] (a2) at (2,0) {};
  \node[node] (a3) at (1,1.732) {}; 

  \draw[edge] (a1) -- (a4);
  \draw[edge] (a4) -- (a2);
  \draw[edge] (a2) -- (a3);
   \draw[edge] (a3) -- (a1);
\end{scope}

\begin{scope}[shift={(5,0)}]
  \node at (-1,2.35) {$(b)$};

  \node[node] (c)  at (0,1) {};
  \node[node] (u)  at (0,2) {};
  \node[node, fill=black] (d)  at (0,0) {};
  \node[node, fill=black] (l)  at (-1,1) {};
  \node[node, fill=black] (r)  at (1,1) {};

  \draw[edge] (c) -- (u);
  \draw[edge] (c) -- (d);
  \draw[edge] (c) -- (l);
  \draw[edge] (c) -- (r);
\end{scope}
\begin{scope}[shift={(10,0)}]
  \node at (-1.5,2.35) {$(c)$};

  \node[node,  fill=black] (e2) at (-1.5,1) {};
  \node[node] (e3) at (-0.5,1) {};
  \node[node,  fill=black] (e4) at (0.5,1) {};

  \draw[edge] (e2) -- (e3);
  \draw[edge] (e3) -- (e4);
 
  \node[node, fill=black] (eup)   at (-0.5,2) {};
  \node[node, fill=black] (edown) at (-0.5,0) {};

  \draw[edge] (e3) -- (eup);
  \draw[edge] (e3) -- (edown);
\end{scope}
\end{tikzpicture}
\caption{Graphic representation of partial simultaneous resolutions. Colored nodes correspond to obstructed 2-cycles. Colored subdiagrams are identified with the subalgebras that compose the Levi subalgebra associated with $\bf\Phi$. Here we show examples of simple flops of length 1, 2, and a non-resolvable $D_4$ singularity. (a) The colouring of the $\tilde{A}_3$ diagram corresponds to the choice of Levi $\mathcal{L}_{\Phi}=A_1^{(1)}\oplus A_1^{(3)} \oplus \langle\alpha_2^*\rangle$. (b) $\tilde{D}_4$ diagram. Coloring corresponds to $\mathcal{L}_{\Phi}=A_1^{(1)}\oplus A_1^{(2)} \oplus  A_1^{(3)}\oplus \langle\alpha_4^*\rangle$. (c) Non-resolvable singularity. Here $\mathcal{L}_{\Phi}=A_1^{(0)}\oplus A_1^{(1)}\oplus A_1^{(2)} \oplus  A_1^{(3)}$.}\label{Fig:ExamplesNonMonFib}
\end{figure}
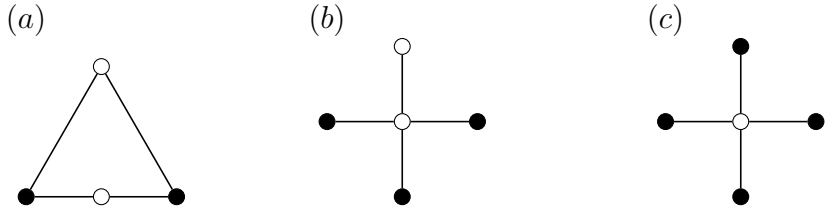

In terms of our Higgs field language, the way we will achieve this is by selecting a \emph{Levi subalgebra} of the semi-simple Lie algebra (see appendix \ref{app:Levi} for two brief definitions). The Higgs field will take non-zero entries along this Levi.
Graphically, this is encoded as follows: Connected subdiagrams of colored nodes are the Dynkin diagrams of the simple summands of the Levi (see Figure~\ref{Fig:ExamplesNonMonFib}).\footnote{
These diagrams involve the exceptional node of the original Dynkin diagram only when the corresponding Levi subalgebra is maximal within the original Lie algebra (or within one of its maximal subalgebras). In this paper, we primarily focus on cases where the coloring does not involve the extended node of the affine Dynkin diagram. This node corresponds to a bound state of the (anti)-fractional branes wrapping the simple roots plus an integral D2-brane. There also exist cDV Calabi–Yau threefolds in which the coloring involves the extended nodes. In this case, the corresponding linear combination of exceptional cycles ceases to define an independent curve in the threefold. Equivalently, this introduces a relation among the simple roots, thereby eliminating one exceptional cycle from the resolution of the threefold.}
These connected subdiagrams are related to subquivers of the original ADE quiver, whose symmetry in their CB is given by the Levi summand.

The distinction between colored and white nodes has an immediate consequence for the probe theory. In the non-monodromic case, each white node corresponds to a fractional brane wrapping a well-defined holomorphic sphere, and the deformation is naturally expressed in terms of the corresponding scalar $\varphi_i$ through the holomorphic-volume formula. In the monodromic case, by contrast, the colored nodes do not correspond to independent holomorphic curves in the threefold. One should therefore not assign separate superpotential deformations to the individual nodes. Rather, each connected colored subdiagram must be treated as a single interacting block.

As we have said, the relevant algebraic structure is the Levi subalgebra selected by the Higgs field $\Phi(w)$. Each connected colored  subquiver has a Coulomb-branch symmetry given by the corresponding summand of the Levi subalgebra. The natural operator carrying this symmetry data is the corresponding moment map $\mu$, which transforms in the adjoint representation of the relevant Levi summand. Although the Higgs field background $\Phi(w)$ takes values in the full ADE Lie algebra, its component along that Levi summand is the one that couples to the corresponding colored block. The canonical gauge-invariant coupling between the geometric data and the probe theory is therefore
\begin{equation}\label{eq:supDefMonodromic}
    \delta W = \kappa \left( \Phi(\varphi_{cm}) , \mu \right)\:.
\end{equation}
The appearance of $\varphi_{\rm cm}=\frac{1}{m}\sum_{i=1}^m\varphi_i$ (where $i$ runs over the colored nodes corresponding to the Levi summand) reflects the fact that, in the presence of monodromy, the corresponding fractional branes are no longer associated with independent motions over the base. Rather, the $\Phi$-background binds them into a single sector, so that the appropriate coordinate entering the deformation is the collective position of the block.
In this sense, \eqref{eq:supDefMonodromic} is the monodromic analogue of the Cartan deformation discussed in Section~\ref{Sec:D2onNonmonodFib}: the individual root volumes are replaced by the adjoint-valued moment map of the colored subquiver, and the Cartan pairing is replaced by the full Killing-form contraction.

The extreme case is when all simple roots are colored, i.e. the Levi subalgebra coincides with the full ADE algebra. When this happens, the resolution of the threefold contains no holomorphic spheres. As a consequence, type IIA on this threefold does not admit fractional branes. The ordinary D2-brane, however, remains, and its $U(1)$ gauge group must be preserved. This $U(1)$ is naturally associated with the extended node of the quiver.

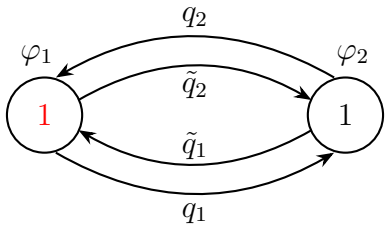
\begin{figure}[H]
\captionsetup{justification=centering}
 \centering
\begin{tikzpicture} [place/.style={circle,draw=black!500,fill=white!20,thick,
inner sep=0pt,minimum size=10mm}, transition/.style={rectangle,draw=black!50,fill=black!20,thick,
inner sep=0pt,minimum size=10mm}, decoration={ markings,
mark=between positions 0.35 and 0.85 step 2mm with {\arrow{stealth}}}]
  \node at ( 2,0) [place] (0) {$1$};  \node at ( 2.1,0.8)  {$\varphi_2$}; 
  \node[red] at (- 2,0) [place] (1) {$1$};  \node at ( -2.1,0.8)  {$\varphi_1$}; 
\draw[thick, -Stealth] (-1.55,0.2)[bend left] to (1.55,0.2); \node at ( 0,1.3)  {$q_2$}; 
\draw[thick, Stealth-] (-1.55,-0.2)[bend right] to (1.55,-0.2);\node at ( 0,-1.3)  {$q_1$};  
\draw[thick, Stealth-] (-1.85,0.5)[bend left] to(1.85,0.5);\node at ( 0,0.4)  {$\tilde{q}_2$}; 
\draw[thick, Stealth-] (1.85,-0.5)[bend left] to(-1.85,-0.5);\node at ( 0,-0.4)  {$\tilde{q}_1$}; 
\end{tikzpicture}
\caption{$A_1$ theory. The $\mathcal{N}=4$ superpotential is given by \ref{eq:N4sup} with $r=1$.}
\label{fig:A1}
\end{figure}

The simplest way to see this mechanism in action is to consider a colored block of type $A_1$, where the moment map can be written explicitly and the induced monopole deformation is completely transparent. First, we put the D2-brane on $A_1\times \mathbb{C}$. The worldvolume theory is the quiver gauge theory in \ref{fig:A1}.

Then, we consider the following Higgs field:
\begin{equation}\label{Eq:C3Phi}
    \Phi = \begin{pmatrix}
    0 & 1 \\ w & 0 \\
    \end{pmatrix} = e_\alpha + w\, e_{-\alpha}
\end{equation}
The corresponding Levi subalgebra is the $A_1$ algebra itself.
The threefold equation is $uv=z^2-w$, i.e. it is $\mathbb{C}^3$. 

According to what we have written above, this $\Phi$ generates the following superpotential deformation
\begin{equation}
\begin{aligned}
    \delta W &= \kappa ( \Phi(\varphi_1) \mu_\alpha ) = w_{-\alpha}+\varphi_1w_\alpha \qquad\mbox{where}\qquad
   \mu_\alpha = \begin{pmatrix}
       \varphi_1 & w_\alpha \\ w_{-\alpha} & -\varphi_1 
   \end{pmatrix} 
\end{aligned}
\end{equation}

To deal with quivers with monopole superpotentials, one follows the prescription in \cite{Collinucci:2016hpz}:
one isolates the abelian node relative to $w_{\pm\alpha}$ by ungauging the adjacent $U(1)$; in our example, this amounts to ungauging the extended node. The isolated theory can then be coupled back to the rest of the quiver by gauging a subgroup of its flavor symmetry.
\begin{figure}[H]
\captionsetup{justification=centering}
 \centering
\begin{tikzpicture} [place/.style={circle,draw=black!500,fill=white!20,thick,
inner sep=0pt,minimum size=10mm}, transition/.style={rectangle,draw=black!50,fill=black!20,thick,
inner sep=0pt,minimum size=10mm}, decoration={ markings,
mark=between positions 0.35 and 0.85 step 2mm with {\arrow{stealth}}}]
  \node at ( 2,0) [place] (0) {$1$};  \node at ( 2.1,0.8)  {$\varphi_2$}; 
  \node[red] at (- 2,0) [place] (1) {$1$};  \node at ( -2.1,0.8)  {$\varphi_1$}; 
\draw[thick, -Stealth] (-1.55,0.2)[bend left] to (1.55,0.2); \node at ( 0,1.3)  {$q_2$}; 
\draw[thick, Stealth-] (-1.55,-0.2)[bend right] to (1.55,-0.2);\node at ( 0,-1.3)  {$q_1$};  
\draw[thick, Stealth-] (-1.85,0.5)[bend left] to(1.85,0.5);\node at ( 0,0.4)  {$\tilde{q}_2$}; 
\draw[thick, Stealth-] (1.85,-0.5)[bend left] to(-1.85,-0.5);\node at ( 0,-0.4)  {$\tilde{q}_1$}; 
\draw[thick, -Stealth] (3,0) to (4,0);
\node[red] at ( 5,0) [place] (2) {$1$}; 
  \node at (9,0) [transition] (3) {$2$};  \node at ( 5.1,0.8)  {$\varphi_1$}; 
\draw[thick] (5.5,0.2) to (8.5,0.2); \node at ( 7,0.45)  {$\tilde{q}_2,q_2$}; 
\draw[thick] (5.5,-0.2) to (8.5,-0.2);\node at ( 7,-0.45)  {$\tilde{q}_1,q_1$}; 

\end{tikzpicture}
\caption{$A_1$ theory. Ungauging of the affine node.}
\end{figure}
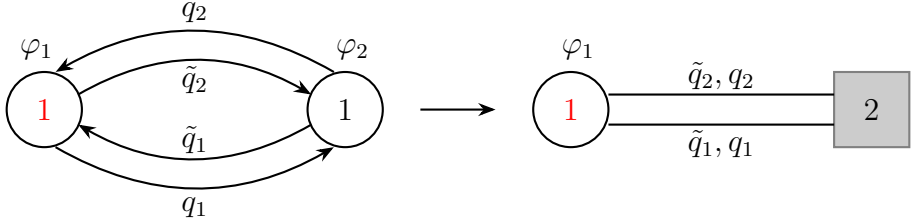

For the isolated (balanced) node, the theory is a 3d $\mathcal{N}=4$ supersymmetric $U(1)$ gauge theory with two flavors, and deformed superpotential
\begin{eqnarray}
    W^{(1)}=\varphi_1(q_1\tilde{q}_1-q_2\tilde{q}_2) + w_{-\alpha}+\varphi_1w_\alpha\:,
\end{eqnarray}
i.e. we obtain the Building Block I in Section~\ref{Sec:MonopoleSuperpot}. The effective theory has no $U(1)$ and a superpotential 
\begin{eqnarray}
    W_{\rm eff}^{(1)}=X_1(Y_1Z_1+\varphi_1-T_1T_2)+\varphi_1(T_1-T_2)\:.
\end{eqnarray}

We now gauge the $U(1)$ of the node 2 back and obtain a $U(1)$ gauge theory with a given number of singlets and superpotential
\begin{eqnarray}
    W_{\rm eff}=X_1(Y_1Z_1+\varphi_1-T_1T_2)+(\varphi_1-\varphi_2)(T_1-T_2)\:.
\end{eqnarray}
One can make a field redefinition $(\varphi_1,\varphi_2,T_1,T_2)\longleftrightarrow (A,B,T_+,T_-)$ in the following way: $A=\varphi_1+Y_1Z_1-T_1T_2$, $B=\varphi_1-\varphi_2$, $T_\pm=T_1\pm T_2$. The superpotential can be then written as
\begin{eqnarray}
    W_{\rm eff}=X_1 A+B\,T_-\:.
\end{eqnarray}
Hence, $X_1,A,B,T_-$ are massive and we can integrate them out. 

The final theory has a quiver with one node and three loops, 
namely the quiver whose moduli space is $\mathbb{C}^3$, in agreement with the $\Phi$ in \eqref{Eq:C3Phi}.  The node corresponds to the D2-brane that probes  $\mathbb{C}^3$ (originally it was the extended node of the $A_1$ affine Dynkin diagram).

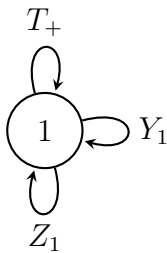
\begin{figure}[H]
\centering
\begin{tikzpicture}[>=stealth, node distance=2cm, thick]

  \node[circle, draw, minimum size=10mm, inner sep=0pt] (v) {$1$};
  \path[->] (v) edge [loop above] node {$T_+$} (v);
  \path[->] (v) edge [loop right] node {$Y_1$} (v);
  \path[->] (v) edge [loop below] node {$Z_1$} (v);
\end{tikzpicture}
\caption{$\mathbb{C}^3$ theory.}
\end{figure}

In the next sections we study interesting monodromic cases.

\subsection{Summary of the algorithm to get the effective 3d $\mathcal{N}=2$ theory}

Given a cDV threefold described by a Higgs field $\Phi(w)$, our prescription is:
\begin{enumerate}
    \item Start from the $\mathcal N=4$ quiver gauge theory on a D2-brane probing the corresponding undeformed ADE surface singularity.
    \item Determine the Levi decomposition encoded by $\Phi(w)$, or equivalently the colored Dynkin diagram specifying which roots are resolved and which ones are affected by monodromy.
    \item For each white node, introduce the polynomial deformation induced by the Cartan component of $\Phi(w)$ through the holomorphic volume formula.
    \item For each connected colored subdiagram, identify the corresponding Coulomb-branch moment map and add the superpotential deformation
    \begin{equation}
        \delta W = \kappa\!\left(\Phi(\varphi_{\rm cm}),\mu\right),
    \end{equation}
    where $\varphi_{\rm cm}$ is the center-of-mass scalar for the corresponding block. We do not currently have a first-principles derivation of why the deformation depends on $\varphi_{\rm cm}$ in precisely this way. Nevertheless, the prescription is strongly supported {\it a posteriori} by the fact that it reproduces the correct effective theories and moduli spaces in all the examples analyzed in this paper.
    \item Use local 3d mirror symmetry, together with ungauging/regauging when needed, to trade the monopole deformations for an effective theory with polynomial superpotential in elementary fields.
    \item Compute the classical moduli space of the resulting $\mathcal N=2$ theory at the origin of the Coulomb branch. In the examples studied in this paper, this reproduces the cDV threefold probed by the D2-brane.
\end{enumerate}

The local mirror-symmetry procedure described above naturally produces an effective superpotential $W_{\rm eff}$ for the infrared theory. Its primary role in our construction is to encode the F-term relations that define the geometric Higgs branch. In particular, $W_{\rm eff}$ will not always be written in the standard form of a quiver potential, namely as a sum of single-trace terms corresponding to concatenations of arrows along closed paths in the quiver.

In the examples studied below, this issue arises in particular for the monodromic description of Reid's pagodas and for simple flops of length two. In those cases, although the superpotential obtained directly from the algorithm is not manifestly a quiver potential, the resulting F-term relations can nevertheless be integrated to a new superpotential $W_{\rm quiv}$ written in terms of admissible path concatenations. The two superpotentials lead to the same F-term relations for the purposes of the moduli-space analysis, and therefore define the same moduli space.

\section{Reid's pagodas}\label{Sec:ReidPagodas}
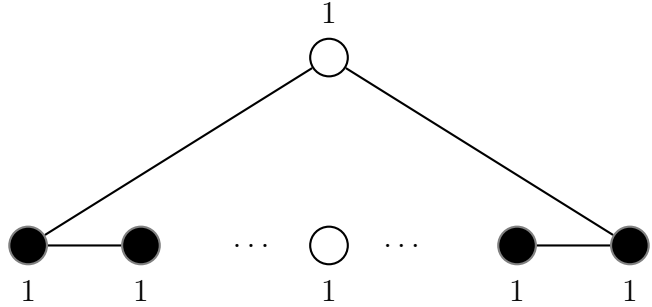
\begin{figure}[H]
\captionsetup{justification=centering}
 \centering
\begin{tikzpicture} [place/.style={circle,draw=black!500,fill=white!,thick,
inner sep=0pt,minimum size=5mm}, transition/.style={circle,draw=black!50,fill=black!200,thick,
inner sep=0pt,minimum size=5mm}, decoration={ markings,
mark=between positions 0.35 and 0.85 step 2mm with {\arrow{stealth}}}]
 \node at ( 2.5,0) [transition] (2) {}; \node at ( 2.5,-0.6) {$1$};
  \node  at (- 2.5,0) [transition] (1) {}; \node at (- 2.5,-0.6) {$1$}; 
 \node  at ( 4,0) [transition] (3) {}; \node at ( 4,-0.6) {$1$};
  \node at ( -4,0) [transition] (4) {}; \node at ( -4,-0.6) {$1$};
   \node at (0,0) [place] (0) {};\node at (0,-0.6) {$1$};
  \node at (0,2.5) [place] (0) {};\node at (0,3.1) {$1$};
  \node at (-1,0) [] {$\dots$};
  \node at (1,0) [] {$\dots$};
  \draw[thick] (4)--(1);
  \draw[thick] (2)--(3);
  \draw[thick] (4)--(0);
  \draw[thick] (0)--(3);
\end{tikzpicture}
\caption{$A_{2k-1}$ partial simultaneous resolution }\label{Fig:DynkPagoda}
\end{figure}

We now return to Reid's Pagoda, studied earlier in Section~\ref{Sec:ReidsPagodaNonMonodromicA1}
as a non-monodromic $A_1$ fibration. Here we re-analyze the same 
geometry from a different perspective: as a monodromic $A_{2k-1}$ 
fibration where most nodes are colored. This provides a highly 
non-trivial check of our method.

The defining equation of the Reid’s Pagoda is
\begin{eqnarray}\label{Eq:equationReidsPagodaSU2k}
    uv=z^{2k}-w^2\:.
\end{eqnarray}
This is the same equation as \eqref{Eq:equationReidsPagoda}, but with the roles of $z$ and $w$ exchanged. The equation \eqref{Eq:equationReidsPagodaSU2k} can be seen as an $A_{2k-1}$ singularity deformed at $w\neq 0$. It is already known that the resolution of this manifold contains a single exceptional sphere. This immediately implies that most of the nodes in the $A_{2k-1}$ Dynkin diagram must be colored, as shown in Fig.~\ref{Fig:DynkPagoda}. The associated Levi subalgebra is
\begin{equation}
\mathcal{L} =  \langle\alpha_{k}^\ast\rangle   \oplus A_{k-1}^{L} \oplus A_{k-1}^{R} \:
\end{equation}
and the Higgs field takes the following block diagonal form
\begin{equation}
    \Phi = \begin{pmatrix}
        \Phi_+ & \\ & \Phi_- \\
    \end{pmatrix}
\end{equation}
with the two blocks given by
\begin{equation}\label{recHiggsU}
\Phi_{\pm} =\left(\begin{array}{cccccc}
0 & 1 & 0 & \cdots && 0 \\
0 & 0 & 1 & \ddots && \vdots\\
\vdots & \ddots & \ddots & \ddots && 0  \\
 0 & \cdots  & 0 & 0 && 1 \\
\pm w & 0 & \cdots & 0 && 0  \\
\end{array} \right)\:,
\end{equation}
i.e.
\begin{equation}
    \Phi(w) = \sum_{m=1}^{k-1} e_{\alpha_m} + w \, e_{-\alpha_1-...-\alpha_{k-1}} \,+\,
     \sum_{m=k+1}^{2k-1} e_{\alpha_m} - w \, e_{-\alpha_{k+1}-\,...\,-\alpha_{2k-1}}\:.
\end{equation}
We notice that $\Phi$ has no component along $\alpha_k^\ast$.

\begin{figure}[t!]\label{qui1}
\captionsetup{justification=centering}
 \centering
\begin{tikzpicture} [place/.style={circle,draw=black!500,fill=white!20,thick,
inner sep=0pt,minimum size=10mm}, transition/.style={rectangle,draw=black!50,fill=black!20,thick,
inner sep=0pt,minimum size=10mm}, decoration={ markings,
mark=between positions 0.35 and 0.85 step 2mm with {\arrow{stealth}}}]
 \node[red] at ( -5,0) [place] (1) {$1$};\node at (-5, -0.8) {$\varphi_1$};\node at (-4, -0.7) {$q_{2}$};\node at (-4, 0.35) {$\tilde{q}_{2}$};
 \node[red] at ( 5,0) [place] (4) {$1$}; \node at (5, -0.8) {$\varphi_{5}$};\node at (4, -0.7) {$q_{5}$};\node at (4, 0.35) {$\tilde{q}_{5}$};
  \node[red] at ( 3,0) [place] (3) {$1$}; \node at (3, -0.8) {$\varphi_4$};\node at (3, 3.4) {$q_{6}$};\node at (3, 2.2) {$\tilde{q}_{6}$};
  \node[red] at (- 3,0) [place] (2) {$1$}; \node at (-3, -0.8) {$\varphi_2$};
  \node at (0,4)[transition] (0) {$1$}; \node at (0, 4.8) {$\varphi_6$};\node at (-3, 3.3) {$q_{1}$};\node at (-3, 2) {$\tilde{q}_{1}$};
   \node at (-1.5,0) (5) {}; \node at (-1.5, -0.8) {$q_{3}$};\node at (-1.5, 0.35) {$\tilde{q}_{3}$};
    \node at (1.5,0) (6) {$ $}; \node at (1.5, -0.8) {$q_{4}$};\node at (1.5, 0.35) {$\tilde{q}_{4}$};
     \node at (0,0) [transition] (7) {$1$}; \node at (0,-0.8) {$\varphi_3$};
 \draw[->, -Stealth, thick] (0)[bend right] to (1);  \draw[->,-Stealth,thick] (1)--(0);\draw[->,-Stealth,thick] (4)[bend right] to (0);\draw[->,-Stealth,thick] (0)--(4);
 \draw[thick, -Stealth] (4)--(3);\draw[thick, -Stealth] (3)[bend right] to(4);
  \draw[thick,Stealth-] (1)--(2);\draw[thick, Stealth-] (2)[bend left] to(1);
  \draw[thick, Stealth-] (2)--(7);\draw[thick, Stealth-] (-0.5, -0.2)[bend left] to(2);
  \draw[thick, -Stealth] (3)--(7);\draw[thick, -Stealth] (0.5, -0.2)[bend right] to(3);
 \draw[thick, red] (-4, 0) ellipse (2 and 1.5);
  \draw[thick, red] (4, 0) ellipse (2 and 1.5);
  \node[red] at (-4, -1.9) {$A_2^{(1)}$};
    \node[red] at (4, -1.9) {$A_2^{(2)}$};
    \draw [arrows = {-Latex[line width=7pt, fill=red, length=15pt]}] (0,-3) -- (0,-4.5);
   \node[red] at ( 5,-6) [place] (9) {$1$}; \node at (5, -6.8) {$\varphi_{5}$};\node at (4, -6.7) {$q_{5}$};\node at (4, -5.65) {$\tilde{q}_{5}$};
  \node[red] at ( 3,-6) [place] (8) {$1$}; \node at (3, -6.8) {$\varphi_4$};\node at (6.7, -5.3) {$q_{6}$};\node at (6.2, -5.25) {$\tilde{q}_{6}$}; 
   \draw[thick, -Stealth] (9)--(8);\draw[thick, -Stealth] (8)[bend right] to(9);
   \node at (6.5, -6) [transition] (10) {$2$};
   \draw[thick] (9)--(10);
    \node at (1.5, -6) [transition] (11) {$2$};
     \draw[thick] (8)--(11);
     \node at (1.7, -5.3) {$q_{4}$};\node at (1.2, -5.25) {$\tilde{q}_{4}$};
 \node[red] at ( -5,-6) [place] (12) {$1$};\node at (-5, -6.8) {$\varphi_1$};\node at (-4, -6.7) {$q_{2}$};\node at (-4, -5.65) {$\tilde{q}_{2}$};     
   \node[red] at (- 3,-6) [place] (13) {$1$}; \node at (-3, -6.8) {$\varphi_2$};    
      \draw[thick,Stealth-] (12)--(13);\draw[thick, Stealth-] (13)[bend left] to(12);
        \node at (-1.5, -6) [transition] (14) {$2$};
        \draw[thick] (13)--(14);
         \node at (-6.5, -6) [transition] (15) {$2$};
          \draw[thick] (12)--(15);
\node at (-1.8, -5.3) {$q_{3}$};\node at (-1.3, -5.25) {$\tilde{q}_{3}$};
\node at (-6.8, -5.3) {$q_{1}$};\node at (-6.3, -5.25) {$\tilde{q}_{1}$};
\end{tikzpicture}
\caption{Example with k=3. Ungauging of the two Levi blocks.}
\end{figure}
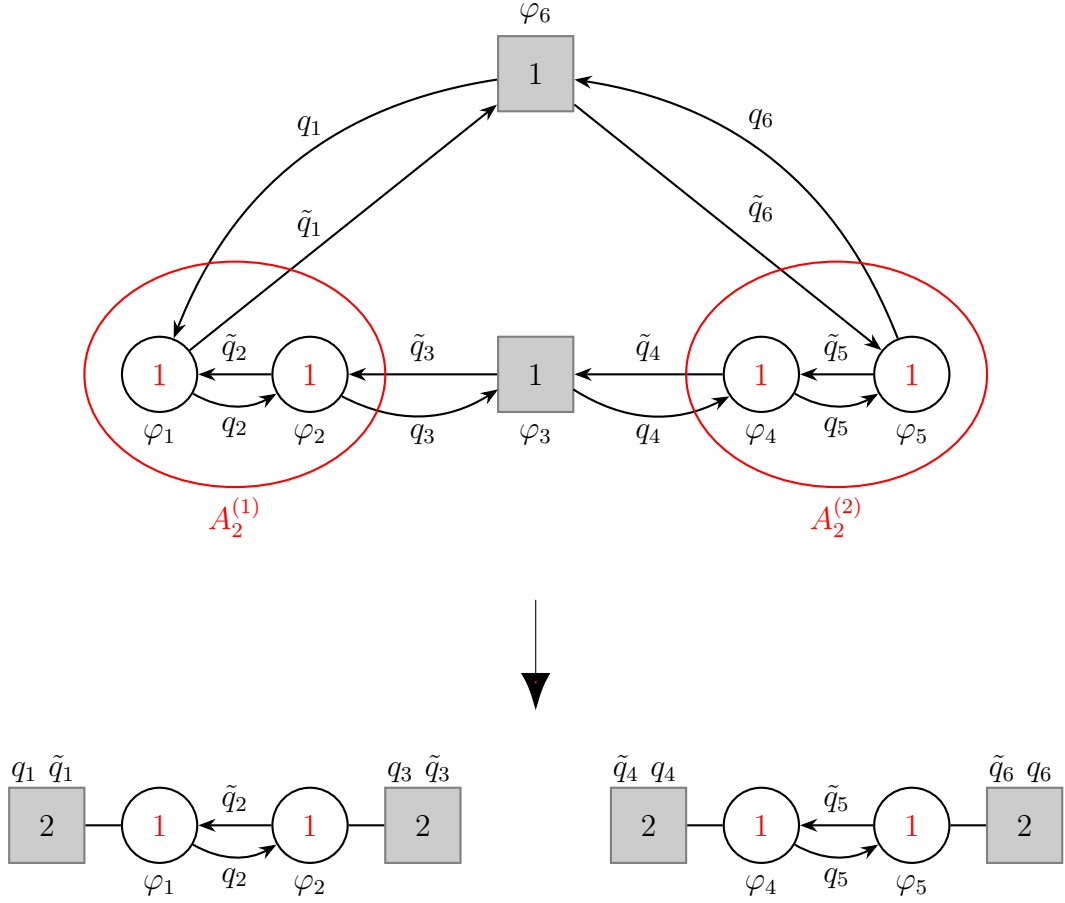

Following the algorithm outlined in Section~\ref{Sec:D2onADEfib}, the superpotential deformation splits into two distinct contributions, each associated with a simple summand of the Levi subalgebra:
\begin{equation}
\begin{aligned}
    \delta W =& \kappa \left( \Phi\left( \tfrac{1}{k-1}\sum_{m=1}^{k-1}\varphi_m\right)\,,\, \mu_L \right) + 
    \kappa \left( \Phi\left( \tfrac{1}{k-1}\sum_{m=k+1}^{2k-1}\varphi_m\right)\,,\, \mu_R \right)\\
    =& \sum_{m=1}^{k-1} w_{-\alpha_m} + \left( \tfrac{1}{k-1}\sum_{m=1}^{k-1}\varphi_m\right) \, w_{+\alpha_1+...+\alpha_{k-1}} \,+\, \\ &+
     \sum_{m=k+1}^{2k-1} w_{-\alpha_m} - \left( \tfrac{1}{k-1}\sum_{m=k+1}^{2k-1}\varphi_m\right) \, w_{+\alpha_{k+1}+...+\alpha_{2k-1}}
\end{aligned}
\end{equation}

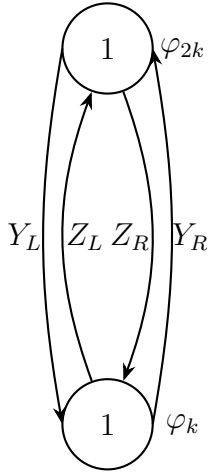
\begin{figure}[t]
\captionsetup{justification=centering}
\centering
\begin{tikzpicture}[
    place/.style={
        circle,
        draw=black,
        fill=white,
        thick,
        inner sep=0pt,
        minimum size=12mm
    }
]

\node[place] (top) at (0,5) {$1$};
\node[place] (bot) at (0,0) {$1$};
\node at (1,5) (2) {$\varphi_{2k}$};
\node at (1,0) (3) {$\varphi_k$};
\node at (-1.1,2.5) {$Y_L$};
\node at (-0.3,2.5) {$Z_L$};
\node at (0.3,2.5) {$Z_R$};
\node at (1.1,2.5) {$Y_R$};

\draw[thick,-{Stealth}] (top) to[bend left=20] (bot);
\draw[thick,{Stealth}-] (top) to[bend right=20] (bot);

\draw[thick,-{Stealth}] (-0.6,5) to[bend right=10] (-0.6,0);
\draw[thick,{Stealth}-] (0.6,5) to[bend left=10] (0.6,0);



\end{tikzpicture}
\caption{Quiver for the effective theory of the $Pagoda_k$.}\label{Fig:effthpartialQuiver}
\end{figure}

We see that each block of $\Phi$ produces a deformation like in Building Block II, see \eqref{Eq:SupN4LinQuivNNode}.
It is then easy to work out the effective potential, by the ungauging/gauging procedure of \cite{Collinucci:2016hpz} and using the Building Block II in Section~\ref{Sec:MonopoleSuperpot}. We obtain a quiver with two nodes (that originally corresponded to the root $\alpha_k$ and the extended node) depicted in Fig.~\ref{Fig:effthpartialQuiver} and superpotential
\begin{equation}\label{Eq:PagodaSuperpotMonodr}
    \begin{aligned}
        W_{\rm eff} = & X_L\left(Y_L Z_L+ \tfrac{1}{k-1}\sum_{m=1}^{k-1}\varphi_m -T_1T_2...T_{k} \right) \\
        + & X_R\left(Y_R Z_R- \tfrac{1}{k-1}\sum_{m=k+1}^{2k-1}\varphi_m -T_{k+1}T_{k+2}...T_{2k} \right) \\ 
        + & \sum_{m=1}^{2k}\varphi_m(T_m-T_{m+1})\:,
    \end{aligned}
\end{equation}
where we have also recoupled the fields $\varphi_k$ and $\varphi_{2k}$ ($T_{2k+1}\equiv T_1$).

Looking at the superpotential \eqref{Eq:PagodaSuperpotMonodr}, we notice several mass terms. In particular the F-terms for $\varphi_i$'s imply the following relations
\begin{equation}
    \left\{\begin{array}{lcl}
     T_m=T_{m+1}-\tfrac{1}{k-1} X_L        &&  \\    
     T_{k+m}=T_{k+m+1}+\tfrac{1}{k-1} X_R   && \\
     T_{k+1} =T_k && \\
     T_{1} = T_{2k} &&\\
    \end{array}\right.
\qquad \rightarrow \qquad
    \left\{\begin{array}{lcl}
     T_m=T_{k}-\tfrac{k-m}{k-1} X_L     && m=1,...,k-1 \\    
     T_{k+m}=T_{2k}+\tfrac{k-m}{k-1} X_R    && m=1,...,k-1 \\
     T_{2k} + X_R = T_k && \\
     T_{k}-X_L = T_{2k} &&\\
    \end{array}\right.
\end{equation}
that can be solved as
\begin{equation}
\begin{aligned}
     T_{k-j}&=\tfrac{k-j-1}{k-1}T_{k}+\tfrac{j}{k-1} T_{2k}    &\qquad& j=1,...,k-1\\
     T_{k+j+1}&=\tfrac{k-j-1}{k-1}T_{k}+\tfrac{j}{k-1} T_{2k}    &\qquad& j=0,...,k-2\\
     X_L&= X_R = T_k-T_{2k} \\
\end{aligned}
\end{equation}

Integrating out the $4k$ massive fields $X_L,X_R,T_1,...,T_{k-1},T_{k+1},...,T_{2k-1},\varphi_1,...,\varphi_{2k}$, we obtain
\begin{equation}\label{Eq:PagodaMonodromicWeff}
\begin{aligned}
    W_{\rm eff} &= (T_k-T_{2k})\left( Y_LZ_L + Y_RZ_R -2\left(\frac{1}{k-1}\right)^k \prod_{j=0}^{k-1}\left[ 
    j\,T_k + (k-j-1)T_{2k}
    \right]  \right) \\
      &= (T_k-T_{2k})  \left( Y_LZ_L + Y_RZ_R - Q_k(T_k,T_{2k}) \right)
\end{aligned}
\end{equation}
where $Q_k(T,T)=2 T^k$.

We are now ready to compute the classical moduli space. The F-term equations are
\begin{eqnarray}
    \begin{array}{lcl}
        \partial_{T_k} W_{\rm eff} &:\qquad& Y_LZ_L + Y_RZ_R - Q_k(T_k,T_{2k}) - (T_k-T_{2k}) \partial_{T_k}Q_k(T_k,T_{2k}) = 0 \\ 
        \partial_{T_{2k}} W_{\rm eff} &:\qquad& Y_LZ_L + Y_RZ_R - Q_k(T_k,T_{2k}) + (T_k-T_{2k}) \partial_{T_{2k}}Q_k(T_k,T_{2k}) = 0 \\
        \partial_{Y_i} W_{\rm eff} &:\qquad& Z_i\,(T_k-T_{2k})=0 \\
        \partial_{Z_i} W_{\rm eff} &:\qquad& Y_i\,(T_k-T_{2k})=0 \\
    \end{array}
\end{eqnarray}
These equations imply $T_k=T_{2k}\equiv T$ and \footnote{The last two equations seems to allow a branch with $Z_i=Y_i=0$; however, the first two homogeneous independent equations in $T_k ,T_{2k}$ would give $T_k=T_{2k}=0$.}
\begin{equation}\label{Eq:PagodakRelEff}
    Y_LZ_L + Y_RZ_R -2 T^k = 0 \:.
\end{equation}
Let us write down the hypersurface equation. The gauge invariants are
\begin{eqnarray}
    A_1\equiv Y_LZ_L\,,\qquad A_2\equiv Y_RZ_R\,,\qquad B_1\equiv Y_LZ_R\,,\qquad B_2\equiv Y_RZ_L\,,
\end{eqnarray}
with relation $B_1B_2=A_1A_2$. Using \eqref{Eq:PagodakRelEff}, one obtains
\begin{eqnarray}
    B_1B_2 = A_1 (- A_1 + 2 T^k)
\end{eqnarray}
By redefining the variables as $B_1=u$, $B_2=v$, $T=z$, $A_1=-w+z^k$ one obtains the defining equation \eqref{Eq:equationReidsPagodaSU2k} of the $k$th Reid's Pagoda

If we consider the F-term equations for the fields $T_i$'s and $X_L,X_R$ in ... we see that on the vacuum the fields $\varphi_i$'s satisfy
\begin{eqnarray}
    \varphi_k=\varphi_{2k}=\tfrac{1}{k-1}\sum_{m=1}^{k-1}\varphi_m = \tfrac{1}{k-1}\sum_{m=k+1}^{2k-1}\varphi_m = w\:.
\end{eqnarray}
We can then correctly conclude that in a generic point of the moduli space, the branes recombine into a normal brane; morever, their location in the $w$-plane is given by the vev for the corresponding fields $\varphi_i$.

\ 

We have derived the quiver and superpotential for a D-brane probing a Reid's Pagoda threefold singularity in two different ways: either by viewing the threefold equation as a non-monodromic family of deformed $A_1$ singularities (Section~\ref{Sec:ReidsPagodaNonMonodromicA1}), or by viewing it as a monodromic family of deformed $A_{2k-1}$ singularities (in the present section).
In both cases, the resulting quiver is the same. Although the corresponding superpotentials take different forms, they generate the same ideal of F-term equations.

In fact, the effective superpotential \eqref{Eq:PagodaMonodromicWeff} is not manifestly of standard quiver-potential form, in the sense that some of its terms are not written as concatenations of arrows along closed paths. However, the corresponding F-term relations can be integrated to an equivalent quiver potential,\footnote{That coincides with \eqref{Eq:PagodaNonMonodromicSuperpot} obtained by considering the Pagoda's threefold as a non-monodromic $A_1$-family.} in the sense discussed at the end of Section~\ref{Sec:D2onADEfib}:
\begin{equation}
    W_{\rm quiv}= (T_k-T_{2k})  \left( Y_LZ_L + Y_RZ_R \right) -\frac{2}{k+1}\left( T_k^{k+1}-T_{2k}^{k+1} \right)\:.
\end{equation}
One can therefore use this equivalent form when interpreting the final theory as a quiver gauge theory.

\section{Simple flops of length 2}\label{Sec:D4families}

In this section we consider the effective theory on the worldvolume of a D2-brane probing threefolds which admit a simple flop of length~2. The singularity has an exceptional locus given by a single $\mathbb{CP}^1$, analogous to the conifold, but---unlike the conifold---it does \emph{not} admit a toric description. A key distinction with the conifold is the length invariant~$\ell$ of the flop, in the sense of Koll\'ar \cite{Kollar1991} and Katz--Morrison \cite{katz1992gorensteinthreefoldsingularitiessmall}, which for simple threefold flops measures the multiplicity of the exceptional $\mathbb{CP}^1$ in the contraction.\footnote{More precisely, if $f:Y\to X$ is the contraction and $C\subset Y$ is the exceptional curve, then the length $\ell$ is the length of the scheme-theoretic fiber $f^{-1}(p)$ at the generic point of $C$, where $p=f(C)$; equivalently, for simple flops it coincides with the multiplicity of $C$ in the scheme-theoretic exceptional fiber.} For the conifold $\ell=1$, while in the present example we have $\ell=2$.

\subsection{The geometry and the Higgs field}

\begin{figure}[h]
\centering 
\begin{tikzpicture}[
  node/.style={circle,draw,inner sep=0pt,minimum size=6pt},
  edge/.style={line width=0.6pt}
]

  \node[node] (c)  at (0,1) {};
  \node[node] (u)  at (0,2) {};
  \node[node, fill=black] (d)  at (0,0) {};
  \node[node, fill=black] (l)  at (-1,1) {};
  \node[node, fill=black] (r)  at (1,1) {};

  \draw[edge] (c) -- (u);
  \draw[edge] (c) -- (d);
  \draw[edge] (c) -- (l);
  \draw[edge] (c) -- (r);
\end{tikzpicture}
\caption{Length 2 flop from the family of deformed $D_4$ singularities. In this construction length 2 flops are three-dimensional cuts in the family of deformed $D_4$ surfaces with small irreducible resolution.}
\label{fig:D4flop2}
\end{figure}
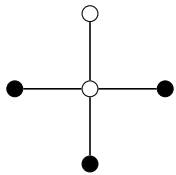

The threefold flops of length 2 are one-parameter $D_4$ families, and hence belong to the class of threefolds we are considering in this paper. They
are described by the colored Dynkin diagram in Figure \ref{fig:D4flop2}. 
The associated Levi subalgebra is 
\begin{equation}
    \mathcal{L} = A_1^{(\alpha_1)}\oplus A_1^{(\alpha_2)} \oplus A_1^{(\alpha_3)} \oplus \langle\alpha_4^\ast \rangle
\end{equation}
The Higgs field is \cite{Collinucci:2022rii,Moleti:2024skd}
\begin{equation}\label{Eq:PhiFl2}
    \Phi = e_{\alpha_1} + \varrho_1  e_{-\alpha_1} +
    e_{\alpha_2} + \varrho_2  e_{-\alpha_2} +
    e_{\alpha_3} + \varrho_3  e_{-\alpha_3} +
    \varrho_4 \alpha_4^\ast\:,
\end{equation}
where $\varrho_i$ are the invariant coordinates of the Levi subalgebra. The threefold with a flop of length 2 is obtained by the choice of a base change $\varrho_i=\varrho_i(w)$, that makes the partial invariant $\varrho_i$ depend on the coordinate $w$.

Written in matrix form, the Higgs field is \cite{Collinucci:2021ofd,Collinucci:2022rii} 
\begin{equation}\label{Flop2Phi}
\Phi(w)=\left(
\begin{array}{cccc|cccc}
\varrho_4  & 1 & 0 & 0 & 0 & 0 & 0 & 0 \\
 \varrho_1  & \varrho_4 & 0 & 0 & 0 & 0 & 0 & 0 \\
 0 & 0 & 0 & 1 & 0 & 0 & 0 & 1 \\
 0 & 0 & \varrho_3 & 0 & 0 & 0 & -1 & 0 \\
 \hline
 0 & 0 & 0 & 0  &  -\varrho_4 & -\varrho_1  & 0 & 0 \\
 0 & 0 & 0 & 0 & -1 & -\varrho_4 & 0 & 0 \\
 0 & 0 & 0 & -\varrho_2  & 0 & 0 & 0 & -\varrho_3  \\
 0 & 0 & \varrho_2 & 0 & 0 & 0 & -1 & 0 \\
\end{array}
\right).
\end{equation}

We now specialize to a class of threefold flops of length two that were studied in \cite{Collinucci:2022rii, vangarderen2022},
and that is obtained by the following base change \cite{Collinucci:2022rii}: 
\begin{equation}\label{Eq:varrhoRuleFl2}
    \varrho_1(w)=c_1 w, \quad \varrho_2(w)=c_2 w, \quad \varrho_3(w)=c_3 w,\quad \varrho_4(w)=c_4 w\:.
\end{equation}
The defining equation can be obtained by plugging \eqref{Eq:PhiFl2} and \eqref{Eq:varrhoRuleFl2} into \eqref{Eq:ADequationsFromPhi}. When we will need a reference example in the family, we will take $c_1=4$, $c_2=c_3=1$ and $c_4=2$; in this case, the defining equation takes the following form:
\begin{equation}\label{Eq:Fl2DefiningEq}
    x^2+z\,y^2 - (z+4w)\left(16z\,w^2+(z+4w-4w^2)^2\right)=0\:.
\end{equation}
By construction it is a one-parameter family of deformed $D_4$ singularities. There are four point-like singularities: at the origin, where the fiber  has a full $D_4$ singularity, and other three points, where the fiber has an $A_1$ singularity.

\subsection{Probing the threefold by a D2-brane}
\begin{figure}[H]
\centering
\begin{tikzpicture} [place/.style={circle,draw=black!500,fill=white!20,thick,
inner sep=0pt,minimum size=10mm}, transition/.style={rectangle,draw=black!50,fill=black!20,thick,
inner sep=0pt,minimum size=10mm}, decoration={ markings,
mark=between positions 0.35 and 0.85 step 2mm with {\arrow{stealth}}}]
\node at (0,0) [place] (1) {$2$};\node at (0,0.8) [] (6) {$\Psi$};
\node at ( 2.5,-2.5) [place] (2) {$1$}; \node at (3.3,-2.5) [] (8) {$\varphi_2$};  \node at ( 1.3,-2.35)  {$\tilde{h}^{(2)}$}; \node at ( 1.3,-0.8)  {$h^{(2)}$}; 
 \node at ( -2.5,-2.5) [place] (3) {$1$};\node at (-3.3,-2.5) [] (9) {$\varphi_1$}; \node at ( -1.2,-2.35)  {$h^{(1)}$}; \node at ( -1.2,-0.8)  {$\tilde{h}^{(1)}$}; 
  \node at ( 2.5,2.5) [place] (4) {$1$};\node at (3.3,2.5) [] (10) {$\varphi_3$};  \node at ( 1.3,2.35)  {$h^{(3)}$}; \node at ( 1.3,0.8)  {$\tilde{h}^{(3)}$}; 
   \node at ( -2.5,2.5) [place] (5) {$1$};\node at (-3.3,2.5) [] (7) {$\varphi_0$}; \node at ( -1.2,2.35)  {$\tilde{h}^{(0)}$}; \node at ( -1.2,0.8)  {$h^{(0)}$}; 
   \draw[->, thick, Stealth-] (1)[bend right] to (2); \draw[->, thick, -Stealth] (1)[bend left] to (2); 
  \draw[->, thick, Stealth-] (1)[bend right] to(3); \draw[->, thick, -Stealth] (1)[bend left] to(3);
  \draw[->, thick, Stealth-] (1)[bend right] to(4); \draw[->, thick, -Stealth] (1)[bend left] to(4);
  \draw[->, thick, Stealth-] (1)[bend right] to(5); \draw[->, thick, -Stealth] (1)[bend left] to(5);
\end{tikzpicture}
\caption{$D_4$ quiver gauge theory.}
\label{fig:quiverd4theory}
\end{figure}
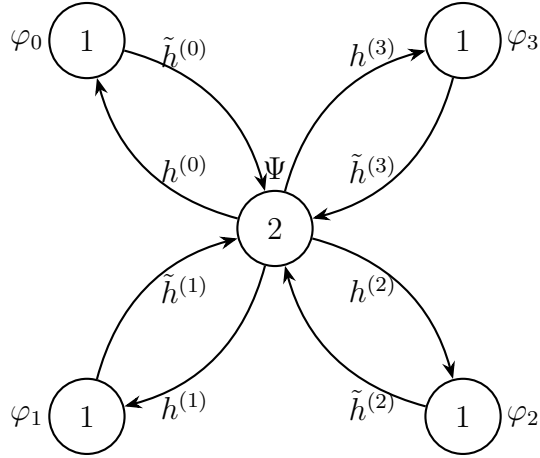
Let us begin from the superpotential of the $D_4$ quiver:
\begin{equation}
W= \sum_{i=0}^3\tr\left[{\Psi}h^{(i)}\tilde{h}^{(i)}\right]-\sum_{i=0}^3 \varphi_i\tilde{h}^{(i)}h^{(i)}, 
\end{equation}
where $(h^{(i)},\tilde{h}^{(i)})$ are bifundamental hypers coupled to the vector multiplets associated to the nodes in \ref{fig:quiverd4theory}. $\Psi$ and the $\varphi_i$'s are the adjoint chirals of the vector multiplets.

Following the algorithm outlined in Section~\ref{Sec:D2onADEfib}, the superpotential deformation generated by the $\Phi(w)$ in \eqref{Eq:PhiFl2} splits into five distinct contributions: three are associated with the three simple summands of the Levi subalgebra, the other two are due to the diagonal $\Phi$-component:
\begin{equation}
\begin{aligned}
    \delta W =&  \delta W_1 + \delta W_2 + \delta W_3 + \delta W_4 +\delta W_0\:,
\end{aligned}
\end{equation}
with
\begin{equation}\label{eq:monopdefflop2}
\begin{aligned}
    \delta W_1 =&  \kappa \left( \Phi\left( \varphi_1\right)\,,\, \mu_1 \right) = w_{-\alpha_1} + \varrho_1(\varphi_1) w_{\alpha_1}\:,\\ 
    \delta W_2 =&  \kappa \left( \Phi\left( \varphi_2\right)\,,\, \mu_2 \right) = w_{-\alpha_2} + \varrho_2(\varphi_2) w_{\alpha_2}\:,\\
    \delta W_3 =&  \kappa \left( \Phi\left( \varphi_3\right)\,,\, \mu_3 \right) = w_{-\alpha_3} + \varrho_3(\varphi_3) w_{\alpha_3}\:,\\
    \delta W_4 =& \tfrac12 \tr \left[\Psi \kappa\left(h_{\alpha_4}, \Phi(\Psi)\right)\right]    
       = \tfrac12 \tr \left[\Psi \varrho_4(\Psi)\right]  \:,\\
    \delta W_0 =& \frac{\varphi_0}{2} \kappa\left(h_{\alpha_0}, \Phi(\varphi_0)\right) 
        = -\varphi_0 \varrho_4(\varphi_0) \:.\\
\end{aligned}
\end{equation}

We now derive the effective superpotential. We ungauge the central node of the quiver. After doing this, each colored node supports an abelian theory with monopole superpotential 
\begin{equation}
    W_i = -\varphi_i(h^{(i)}_1\tilde{h}^{(i)}_1+h^{(i)}_2\tilde{h}^{(i)}_2) + w_{-\alpha_i}+\varrho_i(\varphi_i)w_{\alpha_i}
\end{equation}
Following the computation done in Section~\ref{Sec:MonopoleSuperpot} (Building Block I), the $U(1)$ disappears in the effective theory and we obtain an effective local superpotential
\begin{equation}
\begin{array}{rl}
    W_{{\rm eff},i} =& X_i(Y_i\,Z_i + \varrho_i(\varphi_i) - T^{(i)}_1T^{(i)}_2) - \varphi_i(T^{(i)}_1+T^{(i)}_2)  \\
    =& -X_i \left(\det \mathfrak{M}_i - \varrho_i(\varphi_i) \right) - \varphi_i \tr\mathfrak{M}_i\\
\end{array}
\end{equation}
with $\mathfrak{M}_i \equiv \begin{pmatrix} T^{(i)}_1 & Y_i \\ Z_i & T^{(i)}_2 \end{pmatrix}$.

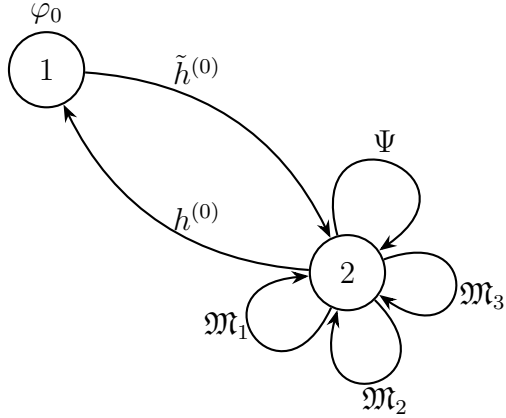
\begin{figure}[H]
\centering
\begin{tikzpicture} [place/.style={circle,draw=black!500,fill=white!20,thick,
inner sep=0pt,minimum size=10mm}, transition/.style={rectangle,draw=black!50,fill=black!20,thick,
inner sep=0pt,minimum size=10mm}, edge/.style={stealth}]
   \node at (8,0) [place] (1) {$2$}; 
     \node at ( 4,2.7) [place] (2) {$1$};\node at (4,3.45) [] (5) {$\varphi_0$};
     \node at (9.8,-0.3) [] (6) {$\mathfrak{M}_3$};  \node at (8.5,-1.7) [] (7) {$\mathfrak{M}_2$}; 
      \node at (6.4,-0.7) [] (8) {$\mathfrak{M}_1$};   \node at (6,2.7) {$\tilde{h}^{(0)}$}; \node at (6,0.75) {$h^{(0)}$};
     \draw[thick] (1) [in=-35,out=20,loop, -Stealth] to (1); 
     \draw[thick] (1) [in=-105,out=-45,loop, -Stealth] to (1); 
     \draw[thick] (1) [in=-175,out=-115,loop, -Stealth] to (1); 
     \draw[thick] (1) [in=35,out=105,loop, -Stealth] to (1); \node at (8.5, 1.75) {$\Psi$};
  \draw[thick, Stealth-]  (1)[bend right] to (2);
   \draw[thick, -Stealth]  (1)[bend left] to (2);
\end{tikzpicture}
\caption{Quiver of the effective $\mathcal{N}=2$ theory that emerges in the IR, when monopole deformations of \ref{eq:monopdefflop2} are turned on.}
\label{fig:flop2IR}
\end{figure}

We now re-gauge the central node symmetry and we obtain the quiver in Figure~\ref{fig:flop2IR} with superpotential
\begin{equation}\label{Eq:EffSupFl2Gen}
    \begin{aligned}
        W_{\rm eff} =& \sum_{i=1}^3\tr\left[{\Psi}\left( \mathfrak{M}_i +h^{(0)}\tilde{h}^{(0)}\right)\right] - \varphi_0\,\tilde{h}^{(0)} h^{(0)}
       \\ 
        &-\sum_{i=1}^3 \left[  X_i \left(\det \mathfrak{M}_i - \varrho_i(\varphi_i) \right)  +\varphi_i \tr\mathfrak{M}_i  \right] +\tfrac{1}{2} \tr \left[\Psi \varrho_4(\Psi)\right]
        -\varphi_0\varrho_4(\varphi_0)  \\
    \end{aligned}
\end{equation}

In the specific example \eqref{Eq:varrhoRuleFl2},  this superpotential becomes

\begin{equation}
    \begin{aligned}
        W_{\rm eff} =& \sum_{i=1}^3\tr\left[{\Psi}\left( \mathfrak{M}_i +h^{(0)}\tilde{h}^{(0)}\right)\right] - \varphi_0\,\tilde{h}^{(0)} h^{(0)}
       \\ 
        &-\sum_{i=1}^3 \left[  X_i \left(\det \mathfrak{M}_i - c_i\varphi_i \right)  +\varphi_i \tr\mathfrak{M}_i  \right] +\tfrac{c_4}{2} \tr \Psi^2-c_4\varphi_0^2 \:. \\
    \end{aligned}
\end{equation}
We can integrate out the massive fields $\varphi_i$ and $X_i$ ($i=1,2,3$), by using their F-term equations:
\begin{equation}\label{Eq:IntegratingOutphiXflop2}
    \begin{aligned}
         \partial_{X_j}W_{\rm eff}=0&: \qquad\varphi_j = -\tfrac{1}{c_j}\,\det\mathfrak{M}_j \\ 
         \partial_{\varphi_j}W_{\rm eff}=0&: \qquad X_j = \tfrac{1}{c_j}\,\tr\mathfrak{M}_j \\ 
    \end{aligned}
\end{equation}
We obtain the effective superpotential:
\begin{equation}\label{Eq:EffSupFl2}
    \begin{aligned}
        W_{\rm eff} =& \sum_{i=1}^3\tr\left[{\Psi}\left( \mathfrak{M}_i +h^{(0)}\tilde{h}^{(0)}\right)\right]  - \varphi_0\,\tilde{h}^{(0)} h^{(0)}
       \\ 
        &+\sum_{i=1}^3 \left[ \tfrac{1}{c_i} \tr \mathfrak{M}_i\, \det \mathfrak{M}_i  \right]  +\tfrac{c_4}{2} \tr \Psi^2-c_4\varphi_0^2 \:,\\
    \end{aligned}
\end{equation}
whose fields are represented in the effective quiver in Figure~\ref{fig:flop2IR}.
We claim that these are the quiver and the superpotential of a D2(D3)-brane probing the threefold we are analysing in this section.

\

We are now ready to work out the moduli space. For convenience, let us first write 
\begin{equation}\label{Eq:MiPsiTracePlusTraceless}
\mathfrak{M}_i= t_i\mathbb{1} + \tilde{\mathfrak{M}}_i \, \mbox{ and }  \, \Psi= \varphi_4 \mathbb{1} + \tilde{\Psi}\qquad \mbox{ with } \quad \tr\tilde{\mathfrak{M}}_i = \tr\tilde{\Psi}=0\:.  
\end{equation}

The relations that one obtains from the superpotential \eqref{Eq:EffSupFl2} are:
\begin{equation}
    \begin{aligned}
         &\partial_{\Psi}W_{\rm eff}=0&:&  \qquad \sum_{i=1}^3\mathfrak{M}_i + h^{(0)}\tilde{h}^{(0)} + c_4\Psi =0\\ 
         &\partial_{\tilde{\mathfrak{M}}_j}W_{\rm eff}=0&:& \qquad \tilde{\Psi} - \tfrac{2t_j}{c_j}\tilde{\mathfrak{M}}_j = 0  \\ 
         &\partial_{t_j}W_{\rm eff}=0&:& \qquad  \varphi_4+\tfrac{1}{c_j} \left( 3 t_i^2 - \tfrac12 \tr \tilde{\mathfrak{M}}_j^2  \right) = 0  \\  
         &\partial_{\varphi_0}W_{\rm eff}=0&:& \qquad \tilde{h}^{(0)} h^{(0)} + 2c_4\varphi_0  \\
         &\partial_{\tilde{h}^{(0)}}W_{\rm eff}=0&:& \qquad \left( \Psi - \varphi_0 \mathbb{1} \right) h^{(0)}   \\
         &\partial_{h^{(0)}}W_{\rm eff}=0&:& \qquad \tilde{h}^{(0)}\left( \Psi - \varphi_0 \mathbb{1} \right) \\         
    \end{aligned}
\end{equation}
Here  we have also used the fact that $\det \tilde{\mathfrak{M}}_i=-\tfrac12 \tr\tilde{\mathfrak{M}}_i^2$.

There are no solutions to these equations, unless\footnote{If all $t_i$'s are non-zero, the $\tilde{\mathfrak{M}}_i$'s are all proportional to $\tilde{\Psi}$ with proportionality factor depending on $t_i$ and $c_i$; hence also $\tr \tilde{\mathfrak{M}}_i^2$ depends on $t_i$ and $c_i$. For generic $c_i$'s one cannot solve both the third and the first relations. The same is true if some $t_i$ is non-zero. Hence, all of them must vanish, and then also $\tilde{\Psi}$ must be equal to zero. }  $\tilde{\Psi}=0$ and $t_i=0$ ($i=1,2,3$). We are then left with 
\begin{equation}
    \sum_{i=1}^3\tilde{\mathfrak{M}}_i + h^{(0)}\tilde{h}^{(0)}+c_4\varphi_4\mathbb{1}=0\,,\qquad
    \varphi_0=\varphi_4=-\frac{\det\tilde{\mathfrak{M}}_j}{c_j} \:. 
\end{equation}

In \cite{Moleti:2024skd} (Section 5.2.3), the authors considered a set of relations formally identical to these. Following their steps, one readily finds that the moduli space of the effective theory coincides with the threefold given by equation~\eqref{Eq:Fl2DefiningEq}. Let us briefly sketch the derivation below.

The moduli space si parametrized by $\varphi_4$ and the gauge invariants build up from $h^{(0)},\tilde{h}^{(0)}$ and the traceless matrices $\tilde{\mathfrak{M}}_i$. One chooses as a basis of gauge invariants
\begin{equation}
A_i\equiv \tr\left[\mathfrak{M}_0\tilde{\mathfrak{M}}_i\right], \qquad B\equiv \tr\left[\mathfrak{M}_0\tilde{\mathfrak{M}}_2\tilde{\mathfrak{M}}_3\right]\:,
\end{equation}
where we have defined $\mathfrak{M}_0\equiv h^{(0)}\tilde{h}^{(0)}$.

The invariants $A_1,\,A_2,\,A_3$ and $B$ are not independent, but $A_1,\,A_2,\,A_3$ satisfy the linear reation
\begin{equation}
\sum_i A_i = -2c_4^2\varphi_4^2 
\end{equation}
and they are also related to $B$ by (see Appendix A in \cite{Moleti:2024skd})
\begin{equation}\label{eq:BA1A2A3equation}
\begin{aligned}
A_1A_2A_3 =& -B^2 +2Bc_4\varphi_4(A_1+\varphi_4(c_4^2\varphi_4+c_1-c_2-c_3) ) \\
&-\varphi_4(c_2A_3+c_3A_2)(A_1+2c_4^2\varphi_4^2)-\varphi_4A_2A_3(c_4^2\varphi_4+c_1)-4c_4^2c_2c_3\varphi_4^4\:.
\end{aligned}
\end{equation}

From now on, we set $c_1=4$, $c_2=c_3=1$ and $c_4=2$ as above.
We write $A_1$ in terms of $A_2$ and $A_3$ by using the first relation, and make the following redefinition of the gauge invariant coordinates: 
\begin{eqnarray}
A_2 &=& \tfrac12\left( y+z -4 w(w-1) \right) \;, \\
A_3 &=& \tfrac12\left( -y+z -4 w(w-1) \right) \;, \\
B &=& -\tfrac12 x +2w(z+2w) \;, \\
\varphi_4 &=& w\;. 
\end{eqnarray}
Plugging these redefinitions into equation~\eqref{eq:BA1A2A3equation}, with the chosen values of the $c_i$, we obtain the defining equation~\eqref{Eq:Fl2DefiningEq} of the threefold under consideration.

\

Also in this case, the superpotential \eqref{Eq:EffSupFl2} obtained directly from the algorithm is not manifestly a standard quiver potential. Nevertheless, its F-term relations admit an equivalent presentation in terms of a quiver potential built from admissible path concatenations. 

In particular, using the identity
\[
\tr \mathfrak{M}\,\det \mathfrak{M}=\frac13\Big((\tr \mathfrak{M})^3-\tr (\mathfrak{M}^3)\Big)
\]
valid for any $2\times2$ matrix, we can replace the non-single-trace term in \eqref{Eq:EffSupFl2} by the cubic quiver-potential term $-\frac{1}{3c_i}\tr(\mathfrak M_i^3)$, up to a term proportional to $(\tr \mathfrak M_i)^3$.
Since, as we have seen above, the F-term equations imply $t_i=\frac12\tr\mathfrak M_i=0$ on the geometric branch, the superpotential \eqref{Eq:EffSupFl2} defines the same moduli-space relations as the superpotential
\begin{equation}
\begin{aligned}
W_{\rm quiv} \;=\;&
\sum_{i=1}^3\tr\!\left[\Psi\left(\mathfrak M_i+h^{(0)}\tilde h^{(0)}\right)\right]
-\varphi_0\,\tilde h^{(0)}h^{(0)}
-\sum_{i=1}^3 \frac{1}{3c_i}\,\tr\!\left(\mathfrak M_i^3\right)
+\frac{c_4}{2}\tr\Psi^2-c_4\varphi_0^2 \; ,
\end{aligned}\nonumber
\end{equation}
 that is single trace. As explained at the end of Section~\ref{Sec:D2onADEfib}, we are therefore free to pass to this equivalent form when describing the infrared quiver theory.

\section{D2-brane probing a non resolvable cDV singularity:  $(A_2,D_4)$ three-fold}\label{Sec:NonResolv}

In this section we will apply our techniques to determine the effective theory of a D2-brane probing threefolds with an isolated singularity that do not admit a crepant resolutions.

\subsection{Geometry and Higgs field $\Phi$}

\begin{figure}[h]
\centering
\begin{tikzpicture}[
  node/.style={circle,draw,inner sep=0pt,minimum size=6pt},
  edge/.style={line width=0.6pt}
]


  \node[node,  fill=black] (e2) at (-1.5,1) {};
  \node[node] (e3) at (-0.5,1) {};
  \node[node,  fill=black] (e4) at (0.5,1) {};

  \draw[edge] (e2) -- (e3);
  \draw[edge] (e3) -- (e4);
 
  \node[node, fill=black] (eup)   at (-0.5,2) {};
  \node[node, fill=black] (edown) at (-0.5,0) {};

  \draw[edge] (e3) -- (eup);
  \draw[edge] (e3) -- (edown);
\end{tikzpicture}
\caption{Non-resolvable singularity. The coloring of the diagram signals the obstruction to the resolution of all the linearly independent 2-cycles.}
\label{fig:nonresolvabled4}
\end{figure}
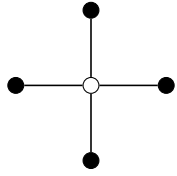

Let us consider the one parameter $D_4$ family, also known as\footnote{Type IIB on this singularity engineers Argyres-Douglas theory of type $(A_2,D_4)$.} $(A_2,D_4)$, given by the equation
\begin{eqnarray}\label{Eq:A2D4equation}
    x^2+z\,y^2+z^3+w^3 = 0
\end{eqnarray}
This threefold was studied in \cite{DeMarco:2021try,DeMarco:2022dgh} with the techniques of \cite{Collinucci:2021wty,Collinucci:2021ofd,Collinucci:2022rii}, so we have the corresponding Higgs field $\Phi$. As explained in \cite{DeMarco:2022dgh}, $\Phi$ belongs to a maximal subalgebra of the Levi subalgebra. In this case the Levi subalgebra is the whole $D_4$ Lie algebra, consistently with the fact that one cannot blow up any two-sphere. If $\Phi$ were a generic element in $\mathcal{L}=D_4$, the threefold would be smooth. Instead in this case, 
\begin{equation}\label{Eq:PhiMforA2D4}
    \Phi \in \mathcal{M}=A_1^{(\alpha_1)}\oplus A_1^{(\alpha_2)} \oplus A_1^{(\alpha_3)} \oplus A_1^{(\alpha_0)} \subset \mathcal{L}=D_4\:,
\end{equation}
where $\alpha_0$ corresponds to the extra node in the extended Dynkin diagram (see Figure~\ref{fig:nonresolvabled4}).

This choice of $\mathcal{M}$ ensures that at $w=0$, the 
singularity is isolated but \emph{non-resolvable}.

The Higgs field in \eqref{Eq:PhiMforA2D4} can be written as
\begin{equation}\label{Eq:PhiFl2nonres}
    \Phi = e_{\alpha_1} + \varrho_1  e_{-\alpha_1} +
    e_{\alpha_2} + \varrho_2  e_{-\alpha_2} +
    e_{\alpha_3} + \varrho_3  e_{-\alpha_3} +
    e_{\alpha_0} + \varrho_0  e_{-\alpha_0}\:,
\end{equation}
where $\varrho_i$ are the invariant coordinates of the Levi subalgebra and $\alpha_0$ is minus the highest root.
Written in matrix form, the Higgs field is \cite{DeMarco:2021try} 
\begin{equation}\label{Flop2Phinonres}
\Phi(w)=\left(
\begin{array}{cccc|cccc}
0  & 1 & 0 & 0 & 0 & \varrho_0 & 0 & 0 \\
 \varrho_1  & 0 & 0 & 0 & -\varrho_0 & 0 & 0 & 0 \\
 0 & 0 & 0 & 1 & 0 & 0 & 0 & 1 \\
 0 & 0 & \varrho_3 & 0 & 0 & 0 & -1 & 0 \\
 \hline
 0 & -1 & 0 & 0  &  0 & -\varrho_1  & 0 & 0 \\
 1 & 0 & 0 & 0 & -1 & 0 & 0 & 0 \\
 0 & 0 & 0 & -\varrho_2  & 0 & 0 & 0 & -\varrho_3  \\
 0 & 0 & \varrho_2 & 0 & 0 & 0 & -1 & 0 \\
\end{array}
\right).
\end{equation}
The particular threefold \eqref{Eq:A2D4equation} is obtained by the base change
\begin{equation}\label{Eq:varrhoRuleFl2nonres}
   \varrho_i=c_i w \quad (i=0,1,2,3) \qquad \mbox{with}\qquad c_1=c_2=c_3=\tfrac14\:, \,\,\, c_0=-\tfrac{3}{4}
    \:.
\end{equation}

\subsection{D2-brane probe theory}

The Higgs field $\Phi(w)$ in \eqref{Flop2Phinonres}, which encodes the background geometry, induces a superpotential deformation on the worldvolume theory of the probe D2-brane. This deformation decomposes into four distinct contributions, each associated with a simple summand of the maximal subalgebra $\mathcal{M} \subset D_4$:
\begin{equation}
\begin{aligned}
    \delta W =&  \delta W_1 + \delta W_2 + \delta W_3 + \delta W_4 
\end{aligned}
\end{equation}
with
\begin{equation}
\begin{aligned}
    \delta W_1 =&  \kappa \left( \Phi\left( \varphi_1\right)\,,\, \mu_1 \right) = w_{-\alpha_1} + \varrho_1(\varphi_1) w_{\alpha_1}\:,\\ 
    \delta W_2 =&  \kappa \left( \Phi\left( \varphi_2\right)\,,\, \mu_2 \right) = w_{-\alpha_2} + \varrho_2(\varphi_2) w_{\alpha_2}\:,\\
    \delta W_3 =&  \kappa \left( \Phi\left( \varphi_3\right)\,,\, \mu_3 \right) = w_{-\alpha_3} + \varrho_3(\varphi_3) w_{\alpha_3}\:,\\
    \delta W_0 =&  \kappa \left( \Phi\left( \varphi_0\right)\,,\, \mu_0 \right) = w_{-\alpha_0} + \varrho_0(\varphi_0) w_{\alpha_0}\:.\\
\end{aligned}
\end{equation}
with $\varrho_i(w)$ given in \eqref{Eq:varrhoRuleFl2nonres}.

We now derive the effective superpotential. We ungauge the central node of the quiver: afterwards, each colored node supports an abelian theory with monopole superpotential 
\begin{equation}
    W_i = -\varphi_i(h^{(i)}_1\tilde{h}^{(i)}_1+h^{(i)}_2\tilde{h}^{(i)}_2) + w_{-\alpha_i}+c_i\varphi_i w_{\alpha_i} \qquad i=0,1,2,3\:.
\end{equation}
Following the computation done in Section~\ref{Sec:MonopoleSuperpot} (Building Block I), the $U(1)$ disappears and we obtain an effective local superpotential
\begin{equation}
\begin{array}{rl}
    W_{{\rm eff},i} =& X_i(Y_i\,Z_i + c_i\varphi_i - T^{(i)}_1T^{(i)}_2) - \varphi_i(T^{(i)}_1+T^{(i)}_2)  \\
    =& -X_i \left(\det \mathfrak{M}_i - \varrho_i(\varphi_i) \right) - \varphi_i \tr\mathfrak{M}_i\\
\end{array}
\end{equation}
with $\mathfrak{M}_i \equiv \begin{pmatrix} T^{(i)}_1 & Y_i \\ Z_i & T^{(i)}_2 \end{pmatrix}$.

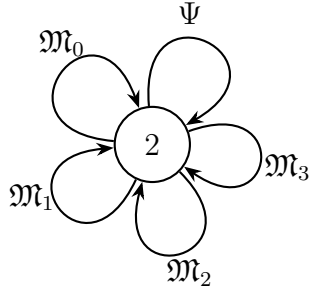
\begin{figure}[H]\label{qui2}
\centering
\begin{tikzpicture} [place/.style={circle,draw=black!500,fill=white!20,thick,
inner sep=0pt,minimum size=10mm}, transition/.style={rectangle,draw=black!50,fill=black!20,thick,
inner sep=0pt,minimum size=10mm}, edge/.style={stealth}]
   \node at (8,0) [place] (1) {$2$}; \node at (6.8,1.35) [] (4) {$\mathfrak{M}_0$};
     \node at (9.8,-0.3) [] (6) {$\mathfrak{M}_3$};  \node at (8.5,-1.7) [] (7) {$\mathfrak{M}_2$}; 
      \node at (6.4,-0.7) [] (8) {$\mathfrak{M}_1$}; 
      \draw[thick] (1) [in=110,out=175,loop, -Stealth] to (1);   
     \draw[thick] (1) [in=-35,out=20,loop, -Stealth] to (1); 
     \draw[thick] (1) [in=-105,out=-45,loop, -Stealth] to (1); 
     \draw[thick] (1) [in=-175,out=-115,loop, -Stealth] to (1); 
     \draw[thick] (1) [in=30,out=95,loop, -Stealth] to (1); \node at (8.5, 1.75) {$\Psi$};
\end{tikzpicture}
\caption{Effective $D_4$ quiver describing the fields coupling to the residual gauge symmetry.}
\label{fig:nonresolvabled4th}
\end{figure}

We now re-gauge the central node symmetry and we obtain the quiver in Figure~\ref{fig:nonresolvabled4th} with superpotential
\begin{equation}\label{Eq:EffSupFl2nonres0}
    \begin{aligned}
        W_{\rm eff} =& \tr\left[\Psi \sum_{i=0}^3 \mathfrak{M}_i\right] + \sum_{i=0}^3 \left(  -X_i \left(\det \mathfrak{M}_i - c_i\varphi_i \right) +(\varphi_4- \varphi_i) \tr\mathfrak{M}_i  \right)  \:.
    \end{aligned}
\end{equation}

Analogously to the flop of length two case, we can integrate out $\varphi_i$ and $X_i$, by using~\eqref{Eq:IntegratingOutphiXflop2}, with now $i=0,1,2,3$. We obtain\footnote{As in the length-two flop case, the effective superpotential can be replaced, for the purposes of the moduli-space analysis, by an equivalent single-trace quiver potential leading to the same F-term relations on the geometric branch.}
\begin{equation}\label{Eq:EffSupFl2nonres}
    \begin{aligned}
        W_{\rm eff} =& \tr\left[{\Psi}\sum_{i=0}^3 \mathfrak{M}_i \right]  
       +\sum_{i=0}^3 \left[ \tfrac{1}{c_i} \tr \mathfrak{M}_i\, \det \mathfrak{M}_i  \right]  \\
       =& \tr\left[\tilde{\Psi}\sum_{i=0}^3 \tilde{\mathfrak{M}}_i \right]+2\varphi_4\sum_{i=0}^3t_i+
       \sum_{i=0}^3\tfrac{t_i}{c_i}\left(t_i^2-\tfrac12\tr\tilde{\mathfrak{M}}_i^2\right)\:, 
    \end{aligned}
\end{equation}
where in the second line we have used \eqref{Eq:MiPsiTracePlusTraceless}.

The relations that one obtains from the superpotential \eqref{Eq:EffSupFl2nonres} are very similar to the simple flop of length two cases studied above:
\begin{equation}\label{Eq:RelationsD4nonres}
    \begin{aligned}
         &\partial_{\tilde{\Psi}}W_{\rm eff}=0&:&  \qquad \sum_{i=0}^3\tilde{\mathfrak{M}}_i  =0\\ 
         &\partial_{\varphi_4}W_{\rm eff}=0&:&  \qquad \sum_{i=0}^3t_i  =0\\ 
         &\partial_{\tilde{\mathfrak{M}}_j}W_{\rm eff}=0&:& \qquad \tilde{\Psi} - \tfrac{2t_j}{c_j}\tilde{\mathfrak{M}}_j = 0  \\ 
         &\partial_{t_j}W_{\rm eff}=0&:& \qquad  \varphi_4+\tfrac{3 t_i^2}{c_j}   - \tfrac{1}{2c_j} \tr \tilde{\mathfrak{M}}_j^2   = 0  \\
    \end{aligned}
\end{equation}

Let us explore the moduli space. The relations \eqref{Eq:RelationsD4nonres} imply $\tilde{\Psi}=0$ and $t_i=0$ ($i=0,...,3$).\footnote{If $\tilde{\Psi}$ and/or some of the $t_i$'s are non-zero, it easy to check that the equations \eqref{Eq:RelationsD4nonres} have no solutions (except for non generic values of $c_i$, that do not include \eqref{Eq:varrhoRuleFl2nonres}).} We are then left with the space parametrized by the four traceless matrices $\mathfrak{M}_i$ and the singlet $\varphi_4$, subject to the relations
\begin{equation}\label{Eq:RelationsA2D4ModSpace}
    \sum_{i=0}^3\tilde{\mathfrak{M}}_i  =0 \qquad\mbox{and}\qquad \tr \tilde{\mathfrak{M}}_j^2 =2c_i\varphi_4\:.
\end{equation}
Out of four traceless $2\times 2$ matrices we can construct the quadratic and the cubic invariants
\begin{equation}
    q_{ij} = \tr \left[ \tilde{\mathfrak{M}}_i\tilde{\mathfrak{M}}_j \right] \qquad\mbox{and}\qquad c_{ijk}=\tr \left[ \tilde{\mathfrak{M}}_i\tilde{\mathfrak{M}}_j\tilde{\mathfrak{M}}_k \right] \:,
\end{equation}
that are related by\footnote{By expanding the traceless matrices $\tilde{\mathfrak{M}}_i$'s in terms of the Pauli matrices, we can associate three-dimensional vectors $v_i$ to them. We then have $q_{ij}=2v_i\cdot v_j$ and $c_{ijk}=2i v_i\cdot(v_j\times v_k)$.}
\begin{equation}
    c_{ijk}^2 = -\frac12 \det \begin{pmatrix}
        q_{ii} & q_{ij} & q_{ik} \\
        q_{ij} & q_{jj} & q_{jk} \\
        q_{ik} & q_{jk} & q_{kk} \\
    \end{pmatrix}
\end{equation}
The quadratic invariants satisfy $q_{ii}=2c_i\varphi_4$ and $\sum_{j=0}^3q_{ij}=0$ $\forall i$. This reduces the number of independent quadratic invariants to two. Let us take $q_{12}$ and $q_{13}$ as the independent ones. The cubic invariants are all proportional to each others.\footnote{Let us see it in terms of the vectors $v_i$'s. Since $\sum_iv_i=0$, we have $v_{i_1}\cdot(v_{i_2}\times v_{i_4}) = v_{i_1}\cdot(v_{i_2}\times (-v_{i_1}-v_{i_2}-v_{i_3}))=
v_{i_1}\cdot(v_{i_2}\times (-v_{i_3}))$. Hence $c_{i_1i_2i_4}=-c_{i_1i_2i_3}=$ for any choice of the ordered quadruple $(i_1,i_2,i_3,i_4)$.} Let us take $c_{123}$ as the independent one. 

We can then are parametrize the moduli space by $\varphi_4,q_{12},q_{13},c_{123}$ that satisfy the relation
\begin{equation}\label{Eq:D4famNonResequation2}
    c_{123}^2 = -\frac12 \det \begin{pmatrix}
        \tfrac{\varphi_4}{2} & q_{12} & q_{13} \\
        q_{12} & \tfrac{\varphi_4}{2} & -q_{12}-q_{13}-\tfrac32\varphi_4 \\
        q_{13} & -q_{12}-q_{13}-\tfrac32\varphi_4 & \tfrac{\varphi_4}{2} \\
    \end{pmatrix}
\end{equation}
where we have used that (with the choice \eqref{Eq:varrhoRuleFl2nonres}) $q_{11}=q_{22}=q_{33}=\tfrac{\varphi_4}{2}$ and $q_{23}=-q_{12}-q_{13}-\tfrac32\varphi_4$.\footnote{This is obtained by taking the proper linear combination of the four relations $\sum_{j=0}^3q_{ij}=0$.}

If we now redefine the coordinates, according to 
\begin{equation}
    c_{123}= \tfrac{1}{2} x  \,,\qquad q_{12} = \tfrac{i}{2} y + \tfrac12 z -  \tfrac12 w \,, \qquad q_{13} = -\tfrac{i}{2}y + \tfrac12 z -  \tfrac12 w\,, \qquad \varphi_4=w\:,
\end{equation}
then the equation \eqref{Eq:D4famNonResequation2} becomes \eqref{Eq:A2D4equation}, confirming once again that the moduli space coincides with the probed threefold.\footnote{We finally observe that using \eqref{Eq:RelationsA2D4ModSpace}, together with \eqref{Eq:IntegratingOutphiXflop2}, one obtains along the moduli space that $\varphi_i=\varphi_4$, for $i=0,1,2,3$. This is consistent with the fact that at generic point in the moduli space, the fractional branes have recombined into a single D2-brane probing smooth points.}

\section{Conclusions}\label{Sec:Conclusions}

In this paper we proposed a systematic probe-brane construction for a broad class of generically non-toric cDV Calabi--Yau threefold singularities. The central input is the Higgs field $\Phi(w)$, which encodes the ADE fibration, its partial simultaneous resolution, and the monodromy of vanishing cycles. Starting from the $\mathcal N=4$ quiver theory on a D2-brane probing the corresponding ADE surface singularity, we explained how $\Phi(w)$ determines a deformation of the probe theory. In non-monodromic cases this deformation is polynomial, while in monodromic cases it involves monopole superpotentials associated with colored Dynkin subdiagrams.

We made extensive use of 3d mirror symmetry, allowing us to replace the monopole deformations by effective polynomial interactions in elementary fields. This yields a practical Lagrangian description of the relevant geometric branch of the moduli space of the probe theory.
One aspect of the derivation deserves further clarification. In the monodromic case, our prescription requires the deformation
$\delta W = \kappa\!\left(\Phi(\varphi_{cm}),\mu\right)$,
where $\varphi_{cm}$ is the center-of-mass scalar of the Dynkin block being collapsed. At present, we do not have a first-principles derivation of why the deformation should depend on $\varphi_{cm}$ precisely in this way. However, the prescription is strongly supported by the fact that it reproduces the expected cDV geometry in several non-trivial examples, including Reid's pagodas, simple flops of length two, and the non-resolvable $(A_2,D_4)$ threefold.

The present analysis should be viewed as a framework rather than as a complete classification. Concretely, our treatment is most explicit for A- and D-type fibrations and for examples that can be reduced to abelian building blocks by mirror-symmetry techniques. Extending the construction to more general non-abelian Levi sectors, and in particular to E-type geometries, remains an important open problem. In fact, this setup complicates both the identification of the appropriate monopole superpotential deformations and the procedure for integrating out monopole operators through mirror symmetry. 

The dictionary developed here between Higgs-field data $\Phi(w)$ and probe-brane Lagrangians offers a useful framework for the study of non-toric Calabi--Yau singularities and their applications in string theory and M-theory.

\section*{Acknowledgments}

We have benefited from fruitful conversations with Federico Carta, Stefano Cremonesi, Mario De Marco, Simone Giacomelli, Amihay Hanany and Andrea Sangiovanni. The research of A.C. is funded through an ARC advanced project, and further supported by IISN-Belgium (convention 4.4503.15).
A.C. is a Senior Research Associate of the F.R.S.-FNRS (Belgium). 
M.M. and R.V. acknowledge support by INFN Iniziativa Specifica ST\&FI.

\appendix
\section{Levi subalgebras}\label{app:Levi}
In this section, we will give two simple working definitions of Levi subalgebras, for reading convenience. The first definition is only valid for $sl_n \mathbb{C}$, the second is general. Both are equivalent, and are consequences of the more formal definition, which we will not give, in terms of parabolics and radicals.

\paragraph{Definition 1: Diagonal Blocks}
A Levi subalgebra of $sl_n \mathbb{C}$ consists in an algebra of block diagonal matrices, i.e. for $sl_5 \mathbb{C}$, one possible Levi subalgebra is given by all matrices of the form:
\begin{equation} \label{eq:blockdiagdef}
    A = \begin{pmatrix}
        \ast & \ast & 0 & 0& 0 \\
        \ast & \ast & 0 & 0 & 0\\
        0 & 0 & \ast & \ast & 0\\
        0 & 0 & \ast & \ast & 0\\
        0 & 0 &0 & 0& \ast
    \end{pmatrix} \in sl_5 \mathbb{C}
\end{equation}

\paragraph{Defnition 2: Centralizers}
Given a \emph{semi-simple} (diagonalizable) element $x \in \mathfrak{g}$ of a Lie algebra, the centralizer $\mathfrak{l} \subset \mathfrak{g}$, defined as
\begin{eqnarray}
    \mathfrak{l}:= \{y \in \mathfrak{g}| [y, x]=0 \}
\end{eqnarray}
is a Levi subalgebra. For instance, in $sl_5 \mathbb{C}$, the Levi subalgebra associated with the following element:
\begin{equation}
    x = \begin{pmatrix}
        m_1 & 0 & 0 & 0 & 0 \\
        0& m_1 & 0 & 0 & 0  \\
        0 & 0 & m_2 & 0 & 0 \\
        0 & 0 & 0 & m_2 & 0 \\
        0 & 0 & 0 & 0 & -2 m_1-2 m_2 \\
    \end{pmatrix}
\end{equation}
consists in the subalgebra defined in the previous definition \eqref{eq:blockdiagdef}.

\providecommand{\href}[2]{#2}\begingroup\raggedright\endgroup

\end{document}